\documentclass[11pt,a4paper]{article}
\usepackage[utf8]{inputenc}
\usepackage[english]{babel}
\usepackage{graphicx}
\usepackage[left=2cm,right=2cm,top=2cm,bottom=2cm]{geometry}

\usepackage{setspace}

\usepackage[electronic]{ifsym}
\usepackage{indentfirst,latexsym}
\usepackage{bm}
\usepackage{bbm}
\usepackage{mathrsfs}
\usepackage{amsmath,amsfonts,amsthm,amscd,amssymb}
\usepackage{pifont}

\usepackage{hyperref}
\usepackage{url}

\usepackage{graphicx}
\usepackage{subfigure}

\usepackage{float}

\usepackage{diagbox}

\usepackage{cite}
\usepackage{cleveref}

\usepackage[all,pdf]{xy}

\usepackage{booktabs}
\usepackage{multirow}
\usepackage{multicol}
\usepackage{stfloats}  
\usepackage{color}

\usepackage{algorithm}         
\usepackage{algorithmicx}
\usepackage{algpseudocode}
\usepackage{float}

\newcommand{\Algr}{\textbf{Algorithm}~}
\newcommand{\Fig}{\textbf{Figure}~}
\newcommand{\Tab}{\textbf{Table}~}
\newcommand{\cpvar}[1]{\texttt{#1}}

\newcommand{\ProcName}[1]{\textsc{#1}}

\algnewcommand\algorithmicswitch{\textbf{switch}}
\algnewcommand\algorithmiccase{\textbf{case}}
\algnewcommand\algorithmicdefault{\textbf{default}}

\algdef{SE}[SWITCH]{Switch}{EndSwitch}[1]{\algorithmicswitch\ #1\ \algorithmicdo}{\algorithmicend\ \algorithmicswitch}%
\algdef{SE}[CASE]{Case}{EndCase}[1]{\algorithmiccase\ #1}{\algorithmicend\ \algorithmiccase}%
\algdef{SE}[DEFAULT]{Default}{EndDefault}[1]{\algorithmicdefault\ #1}{\algorithmicend\ \algorithmicdefault}%

\makeatletter
\renewcommand{\ALG@name}{Algorithm}
\newenvironment{breakablealgorithm}
{% begin of the breakablealgorithm
	\begin{center}
		\refstepcounter{algorithm}% New algorithm
		\setlength{\baselineskip}{15pt} 
		\renewcommand{\caption}[2][\relax]{% Make a new \caption
			\hrule height.9pt depth0pt \kern3pt
			{\raggedright\textbf{\ALG@name~\thealgorithm} ##2\par}%
			\ifx\relax##1\relax % #1 is \relax
			\addcontentsline{loa}{algorithm}{
				\protect\numberline{\thealgorithm}##2}%
			\else % #1 is not \relax
			\addcontentsline{loa}{algorithm}{
				\protect\numberline{\thealgorithm}##1}%
			\fi
			\kern2pt\hrule\kern2pt
		}
	}{% end of the breakablealgorithm
		\kern3pt\hrule\relax% \@fs@post for \@fs@ruled 
	\end{center}
}

\usepackage{listings}

\DeclareMathOperator{\dif}{d}  

\DeclareMathOperator{\Var}{Var}
\DeclareMathOperator{\Std}{Std}

\DeclareMathOperator{\FFT}{FFT}

\newcommand{\RRS}[1]{\mathscr{R}_{#1}}

\newcommand{\scru}[2]{{#1}^{\mathrm{#2}}}
\newcommand{\scrd}[2]{{#1}_{\mathrm{#2}}}

\renewcommand{\vec}[1]{\bm{#1}}
\newcommand{\mean}[1]{\overline{#1}}
\newcommand{\card}[1]{\left| #1 \right|}
\newcommand{\mfloor}[1]{ \left\lfloor {#1} \right\rfloor }
\newcommand{\mceil}[1]{ \left\lceil {#1} \right\rceil }
\newcommand{\mpair}[2]{ \left\langle {#1}, {#2} \right\rangle}

\newcommand{\mat}[1]{\bm{#1}}

\newcommand{\mi} {\mathrm{i}}

\newcommand{\me} {\mathrm{e}}
\newcommand{\set}[1]{\left\{ #1 \right\}}
\newcommand{\seq}[1]{\langle #1 \rangle}
\newcommand{\abs}[1]{\left| #1 \right|}

\newcommand{\normp}[2]{{\left\lVert #1 \right\rVert}_{#2}}

\newcommand{\trsp}[1]{{#1}^\textsf{T}}

\newcommand{\inv}[1]{#1^{-1}}
\newcommand{\ES}[3]{\mathbb{#1}^{{#2}\times {#3}}}     % Euclidean space

\DeclareMathOperator{\E}{E}
\newcommand{\Expt}[1]{\E\set{#1}}     % Expectation operator
\DeclareMathOperator{\A}{\mathcal{A}}
\newcommand{\Ave}[4]{\A_{#1}^{#2:#3}\set{#4}}     % Expectation operator
\newcommand{\OpStd}[4]{\mathcal{S}_{#1}^{#2:#3}\set{#4}}  % Standard deviation operator
\newcommand{\OpCsum}[4]{\mathcal{C}_{#1}^{#2:#3}\set{#4}} % cumulative sum

\newtheorem{thm}{Theorem}

\newtheorem{defn}[thm]{Definition}

\author{Hong-Yan Zhang$^\dag$\thanks{Correspondence author, email: hongyan@hainnu.edu.cn}, Zhi-Qiang Feng$^\dag$, 
Si-Yu Feng$^\ddag$ and Yu Zhou$^\dag$\\
$\dag$ \small{\textit{School of Information Science and Technology, Hainan Normal University, Haikou 571158, China}}\\
$\ddag$ \small{\textit{School of  Geography, University of Leeds, LS2 9JT, United Kingdom}}}
\title{\textbf{Typical Algorithms for
Estimating Hurst Exponent: A Data Analyst's Perspective}}
\begin{document}
\maketitle

\begin{abstract}
The Hurst exponent is a significant metric for characterizing time sequences  with long-term memory property and it  arises in many fields such as physics, engineering, mathematics, statistics, economics, psychology, and so on. The available methods for estimating the Hurst exponent can be categorized into time-domain and spectrum-domain methods based on the representation of the time sequence, and into linear regression and Bayesian method based on parameter estimation techniques. Although there are various estimation methods for the Hurst exponent, there are still some disadvantages that should be overcome: firstly, the estimation methods are mathematics-oriented instead of engineering-oriented; secondly, the accuracy and effectiveness of the estimation algorithms are inadequately assessed; thirdly, the framework of classification  for the estimation methods are insufficient; and lastly there is a lack of clear guidance for selecting proper estimation in practical problems involved in data analysis.
The contributions of this paper lie in four aspects: 1) the optimal sequence partition method is proposed for designing the estimation algorithms for Hurst exponent; 2) the algorithmic pseudo-codes  are adopted to describe the estimation algorithms, which improves the understandability and usability of the estimation methods and also reduces the difficulty of implementation  with computer programming languages; 3) the performance assessment is carried for the typical estimation algorithms via the ideal time sequence with given Hurst exponent and the practical time sequence captured in applications;
4) the guidance for selecting proper algorithms  for estimating the Hurst exponent is presented and discussed. It is expected that the systematic survey of available estimation algorithms could help the users  to understand the principles and the assessment of the various estimation methods could help the users to select, implement and apply the estimation algorithms of interest in practical situations in an easy way.
\end{abstract}

\tableofcontents

\section{Introduction}\label{S:1}

The \textit{long-term memory} (LTM) of \textit{time sequence} (TS)  was discovered in 1951 by Harold E. Hurst\cite{Hurst1951long, Hurst1965long}. The LTM is also named with the \textit{long range dependence} (LRD). The LTM exists in a wide range of natural phenomena, such as rainfall, tree rings, solar flares and so on \cite{Hurst1956methods}. In order to qualitatively explore the the changes of river water level, Hurst proposed the exponent, denoted by $H$ and named with \textit{Hurst exponent} or \textit{Hurst index} for his contribution, to characterize the \textit{time sequence with long-term memory} (TS-LTM). 

\begin{figure*}[htbp]
	\centering
	\includegraphics[width=\textwidth]{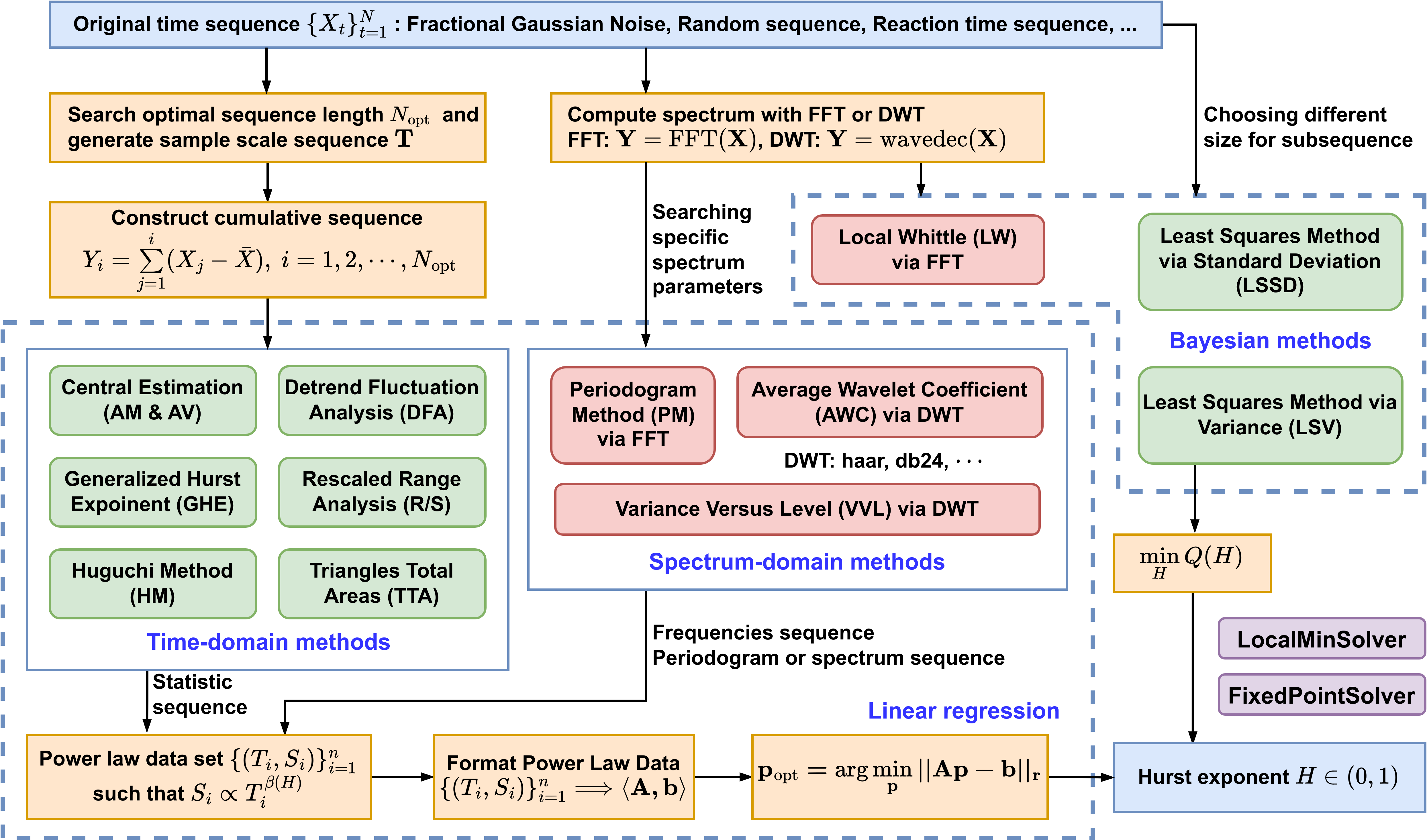}
	\caption{Typical Algorithms and Technical Framework for Estimating Hurst Exponent of TS-LTM}
	\label{fig-flow}
\end{figure*}

It is a fundamental problem to estimate the Hurst exponent for the time sequence with LTM. In the past decades, researchers developed various  methods for estimating the Hurst exponent.
\Fig \ref{fig-flow} illustrates $13$ typical methods and technical framework for estimating the Hurst exponent of TS-LTM.
The estimation methods can be classified with different categories based on two criteria:
\begin{itemize}
	\item[\ding{172}] time-domain methods and spectrum-domain methods based on the representation of TS;
	\item[\ding{173}] linear regression methods and Bayesian methods based on the parameter estimation method.
\end{itemize}

As listed in \Tab \ref{table-method}, there are four types of estimation methods according to the representation ways of TS and the parameter estimation techniques for the Hurst exponent. We now give a brief introduction to the history and ideas behind them.

\begin{table*}[htb]
\caption{Classification of 13 typical estimation methods}\label{table-method}
\begin{tabular}{|l|ll|}
\hline
\diagbox{\textsc{Sequence Representation}}{\rotatebox{-18}{\textsc{Method}}}{\textsc{Parameter Estimation}}		& \textbf{Linear Regression}     & \textbf{Bayesian} \\
\hline
		\textbf{Time-domain}      & AM, AV, GHE, HM,   R/S, DFA, TTA   & LSSD, LSV       \\
		\textbf{Spectrum-domain} &  PM, AWC, VVL                  & LW  \\
\hline
\end{tabular}
\end{table*}

The time-domain estimation method has attracted lots of researchers for several decades. For instance, Beran \cite{Beran1994statistics} constructed the \textit{aggregate variance } (AV) method  for estimating the Hurst exponent using the sample variance  in 1994. After that, Taqqu \cite{Taqqu1995estimators} proposed the \textit{absolute moments } (AM) method  in 1995. The AV method and AM method are unified as the \textit{central estimation}. In 1991, Barab\'{a}si \cite{Barabasi1991multifractality} proposed the \textit{generalized Hurst exponent} (GHE) method via the $q$-th order moments for estimating the Hurst exponent, which is similar to the central estimation. In 1988, the 
\textit{Higuchi method} \cite{Higuchi1988approach} was constructed to calculate the fractal dimension of a time sequence.  Due to the constant offset between the fractal dimension and the Hurst exponent, this method can be applied to estimate the Hurst exponent of a time sequence. In 1994, Peng et al. \cite{Peng1994mosaic} proposed the \textit{detrended fluctuation analysis} (DFA) method  which was also called \textit{residuals of regression} method  \cite{Taqqu1995estimators}. The DFA method was initially utilized to analyze whether non-coding regions in DNA sequences have long-range correlations, and the resulting value is the Hurst exponent. Inspired by the DFA method, Lotfalinezhad et al.\cite{Lotfalinezhad2020tta}  and proposed the \textit{triangles total areas} (TTA) method in 2020 to estimate the Hurst exponent via triangles total areas. Subsequently, G\'{o}mez et al. \cite{Gomez2021theoretical} an effective theoretical framework for the TTA method  and a slightly different method called \textit{triangle area} (TA) method in 2021. In addition to these conventional time-domain methods, Tyralis \cite{Tyralis2011simultaneous} and Koutsoyiannis \cite{Koutsoyiannis2003climate} have proposed two Bayesian statistics estimation methods --- \textit{least squares via standard deviation} (LSSD) and  \textit{least squares via variance} (LSV) --- based on serial variance and standard deviation in 2003 respectively. In 2023, Likens \cite{Likens2023better} compared the performances of these two Bayesian methods and the DFA method and declared that the Bayesian methods are better than the DFA method, especially when the length of the TS is short.

The spectrum-domain estimators have been studied since the beginning of the era of 1980s. For example, Hosking et al. \cite{Hosking1981} discovered that certain time sequences have similar spectral density features in the low-frequency range in 1981. Subsequently, this property was used by Geweke and Porter-Hudak \cite{Geweke1983estimation} to propose the \textit{periodogram} (PM) method  for estimating the Hurst exponent. Moreover, this property was also used by K{\"u}nsch   \cite{Kunsch2020statistical} and Robinson  \cite{Robinson1995LocalWhittle} to develop the \textit{local whittle } (LW) method which is also named \textit{Gaussian semi-parametric estimation}. Phillip systematically assessed these two spectrum-domain methods in 2004 \cite{Phillips2004local}.  In 1989, Flandrin \cite{Flandrin1989spectrum} discovered the relation of wavelet transform 
to fractional Brownian motion. In 1992, Flandrin found \cite{Flandrin1992wavelet} the relationship between the variance of wavelet coefficients and the Hurst exponent based on the discrete wavelet transform . In 1998, Ingve Simonsen et al. \cite{Simonsen1998determination} proposed the \textit{average wavelet coefficient method} (AWC)   for estimating the Hurst exponent using the DWT, and then applied this method to measure the anti-correlation in the Nordic electricity spot market in 2003  \cite{Simonsen2003measuring}.

In recent years, the emphasis of the researchers are focused on the applications of Hurst exponent and performance evaluation of the estimation methods. We now give some examples here.
Zhang et al. \cite{Zhang2017r} in 2017 analyzed the reaction time data   and Li et al.\cite{Li2018systematic} in 2018 studied the features clustering problem with the R/S analysis. 
Moumdjian et al. \cite{Moumdjian2020detrended} applied the DFA method to calculate the fractal statistical properties of the gait time-series to quantify gait dynamics by the outcome measure alpha in 2020.
Shang \cite{Shang2020comparison} in 2020 compared the limited sample variance, variance and mean square deviation of some Hurst exponent estimators in time and spectrum domain with the ARFIMA model. 
 Hamza et al. \cite{Hamza2021comparison} in 2021 proposed the \textit{mean absolute scaled error} (MASE) index of different Hurst exponent estimators based on the \textit{Fractional Brownian motion} (FBM).  
Lahmiri \cite{Lahmiri2023100142} in 2023 used the GHE method to estimate the Hurst exponent of electrocardiogram data, providing auxiliary features for the classification of heart disease data.
Recently, Paul et al. \cite{Paul2024} in 2024 has taken the Hurst exponent to measure and compare the statistical properties of fracture profiles under various micro-structural conditions, correlating these characteristics with the fracture mechanisms of the material.

Although various methods are discussed in literature, there are still some deficiencies from the perspective of data analysts: 
\begin{itemize}
\item the descriptions of the estimation methods are just mathematics-oriented and the general algorithmic pseudo-codes are missing; 
\item the effectiveness and accuracy of the estimation algorithms are not clear; 
\item the classification of estimation methods is not considered; 
\item the technical framework for estimating the Hurst exponent is missing; and
\item a lack of guidance for selecting the estimation methods decreases the  efficiency of data analysis.
\end{itemize}
Motivated by the understandability of the estimation principle and the usability of estimation algorithms for the Hurst exponent arising in data analysis, our contributions of this work lie in the following aspects:
\begin{itemize}
    \item[1)] the optimal sequence partition method is proposed for designing the estimation algorithms for Hurst exponent;
	\item[2)] the mathematical computational methods for estimating the Hurst exponent are converted to algorithmic pseudo-codes in the sense of computer science, which improves the understandability and usability of the estimation methods and also reduces the difficulty of implementation these methods with concrete  high-level computer programming languages such as C/C++, Octave/MATLAB, SciLab, Python, Julia, R, Java, Rust and so on; 
	\item[3)] the performance assessment is carried for the typical estimation algorithms via the ideal time sequences with given Hurst exponent and the practical data of TS-LTM;
	\item[4)]  guidance for selecting the algorithms  for estimating the Hurst exponent is presented based on the performance evaluation.    
\end{itemize}

The contents of this paper are organized as follows: Section \ref{sec-preliminaries} provides the preliminaries for the data analysts; 
Section \ref{sec-opt-seq-partition} deals with the optimal sequence partition; Section \ref{sec-typical-methods-for-H} copes with the typical methods for estimating the Hurst exponent; Section \ref{sec-V-and-V} is devoted to the performance assessment for the typical algorithms; and Section \ref{sec-conclusions} summarizes the conclusions. For the convenience of reading for the readers, the abbreviations used in this paper has been summarized in \Tab \ref{tab-abbreviation}.
\begin{table*}[htbp]
\caption{Nomenclatures}\label{tab-abbreviation}
\begin{tabular}{clp{0.75\textwidth}}
	\toprule
	\textbf{No.} & \textbf{Abbreviation} & \textbf{Interpretation} \\
	\midrule\midrule
	1 & TS & Time Sequence\\
	2 & LTM, LRD  & Long-Term Memory,  also named with the Long Range Dependence \\
	3 & SRD & Short Range Dependence \\
	4 & TS-LTM & Time Sequence with Long-Term Memory \\
	5 & FGN & Fractional Gaussian Noise, which is  used as sample sequence for verification and validation\\
	6 & MASE & Mean Absolute Scaled Error\\
	7 & FBM &  Fractional Brownian Motion \\ 
	8 & DFT, IDFT & Discrete Fourier Transform, Inverse Discrete Fourier Transform \\
	9 & FFT & Fast Fourier Transform \\
	10 & DWT & Discrete Wavelet Transform \\ 	
	11 & LS  & Least Squares \\
	12 & HS  & Human-Speaker \\
	13 & TTS & Text-To-Speech \\
	14 & RT  & Reaction Time \\
	15 & MLE & Maximum Likelihood Estimation\\
	16 & KS  & Kolmogrov-Smirnov \\
	\midrule
	M01 & AM & Absolute Moments, where the first order central moment is used in Central estimation\\
	M02 & AV & Aggregate Variances, where the second order central moment is used in Central estimation\\
	M03 & GHE & Generalized Hurst Exponent $H(q)$, where $q$ is the order\\
	M04 & HM & Higuchi method, which was constructed by Higuchi in 1988\\
	M05 & DFA & Detrended Fluctuation Analysis, which used the residual of regression\\
	M06 & R/S    & Rescaled Range Analysis, where $R=\max\limits_{t} X_t  - \min\limits_{t} X_t$ and $S = \Std(X_t) $ for the TS $\set{X_t}^N_{t=1}$  \\ 
	M07 & TTA & Triangles Total Areas\\
	M08 & PM & Periodogram Method \\
	M09 & AWC & Average Wavelet Coefficient\\
	M10 & VVL & Variance Versus Level\\
	M11 & LW & Local Whittle, also named Gauss semi-parametric estimation\\
	M12& LSSD & Least Squares via Standard Deviation\\
	M13 & LSV & Least Squares via Variance\\
	\bottomrule
\end{tabular}
\end{table*}

\section{Preliminaries} \label{sec-preliminaries}

\subsection{Hurst Exponent}

Mathematically, a time sequence  is just a stochastic process with a discrete parameter ``time'', which can be denoted by $X: \Omega\times \mathbb{Z} \to \mathbb{R}, (\omega, t) \mapsto x$ or simply by
\begin{equation} \label{eq-Xt-def}
x = X(\omega,t) = X_t(\omega)
\end{equation}
where the $\Omega$ is the sample space of a probability space $(\Omega, \mathscr{F}, \Pr)$ and the $\omega\in\Omega$ in \eqref{eq-Xt-def} is a sample point. The mapping $X(\omega,t) = X_t(\omega)$ in \eqref{eq-Xt-def} could be denoted by $X_t$ by omitting the variable $\omega\in \Omega$  according to the custom of data analysis and signals processing.   
The Hurst exponent is closely related to the stationary TS-LTM \cite{Beran1994statistics}:
\begin{defn}
Let $ X_t$ be a stationary stochastic process and $k\in \mathbb{N}$ be the time lagging parameter, $ \rho(k)$  is the autocorrelation function of $X_t$. If there exists a real number $ \alpha\in(0,1) $ and a constant $ c_\rho>0 $ such that asymptotically $\rho(k)\sim c_\rho k^{\alpha}$ for $k\to \infty$, viz.
	\begin{equation}\label{LRD}
	\lim\limits_{k\rightarrow\infty}\dfrac{\rho(k)}{c_\rho k^{-\alpha}}=1,
	\end{equation}
then $X_t$ is called a stationary process with long-term memory.
\end{defn}
There are various jargons for the  LTM of a stationary process such as \textit{long-range dependence},  \textit{strong dependence}, \textit{slowly decaying} and \textit{long-range correlations}.
The LTM is the counterpart of \textit{short range dependency} (SRD). The autocorrelation function of a time sequence with short-range dependency exponentially decay to zero when the time lag increases. In contrast, the autocorrelation function of a time sequence with long-term memory decays moderately, following a power-law decay pattern as described by \eqref{LRD}.

The Hurst exponent $H$ is a measure of the long-term memory capability of a time sequence. Usually, the range of $H$ is the open interval $(0,1)$   \cite{Beran1994statistics}. A value of $H\in(0.5,1)$ indicates the LTM or LRD while $H\in(0,0.5)$ indicates the SRD, and $H=0.5$ represents a random sequence in which observations are completely uncorrelated. The essence of the Hurst exponent is to reflect how the range of fluctuation in a TS changes with the time span.  

Let $\Expt{\cdot}$ be the expectation operator and 
$\Ave{t}{k}{m}{\cdot}$ be the arithmetic average operator  defined by
 \begin{equation}\label{average-operator}
\Ave{t}{k}{m}{X_{\alpha t + \beta s}} = \frac{1}{m-k+1}\sum^m_{t=k}X_{\alpha t +\beta s}
\end{equation}
where the $t$ in \eqref{average-operator} denotes the variable for summation, $k$ and $m$ denote the range of $t$. 
Particularly, for $(\alpha, \beta, m, k) = (1, 0, 1, n)$, the notation for the arithmetic average can be simplified by
\begin{equation} \label{eq-Xt-mean}
\mean{X} = \A(X_t) = \Ave{t}{1}{n}{X_t}=\frac{1}{n}\sum^n_{t=1}X_t.
\end{equation} 
Let $\OpStd{t}{1}{n}{\cdot}$ be the standard deviation operator, we have
\begin{equation}\label{std-operator}
\OpStd{t}{1}{n}{X_t} = \sqrt{\frac{1}{n-1}\sum^n_{t=1} (X_t - \A(X_t))^2} 
\end{equation}
Equations \eqref{eq-Xt-mean} and \eqref{std-operator} are fundamental for statistics computing for time sequences. For the $n$ consecutive samples $\set{X_t}^n_{t=1}$, the range and standard deviation are specified by
\begin{equation} \label{def-Rn}
R(n) = \max_{1\le t\le n} X_t - \min_{1\le t\le n} X_t
\end{equation}
and 
\begin{equation} \label{def-Sn}
S(n) = \OpStd{t}{1}{n}{X_t}
\end{equation}
respectively. The R/S statistic is defined by the ratio of $R(n)$ in \eqref{def-Rn} to $S(n)$ in \eqref{def-Sn}, viz.
\begin{equation} \label{def-RRS}
\RRS{X}(n) = \frac{R(n)}{S(n)}.
\end{equation}
Historically, it was Hurst who discovered that there exists a constant $H$ such that  the $\RRS{X}(n)$ in \eqref{def-RRS} is dominated by the power $n^H$ asymptotically. 
It is a  fundamental problem that how to estimate the Hurst exponent of TS robustly, precisely and efficiently. Currently, there are various methods for estimating the Hurst exponent. According to the equivalent representation domain for the time sequence, there are two fundamental categories for the available estimation methods: 
\begin{itemize}
\item time-domain method, in which the TS $\set{X_t: 1\le t\le N}$ is used directly and the Hurst exponent is estimated by revealing the characteristics of some statistical properties change with the variation of observation period. 
\item spectrum-domain method, in which the \textit{discrete Fourier transform} (DFT) and \textit{discrete wavelet transform} (DWT) are taken for the time sequence $\set{X_t: 1\le t\le N}$.
\end{itemize}

Mathematically, Hurst's observation
can be expressed by
\begin{equation}\label{Intro-1}
\Expt{\RRS{X}(n)} \sim n^H 
\end{equation}
asymptotically for sufficiently large $n$, in which $\E$ is the operator of expectation in the sense of probability and mathematical statistics. 
The method of estimating the Hurst exponent $H$ according to \eqref{Intro-1} is called the
\textit{R/S analysis} or \textit{range rescaled analysis} since $R(n)/S(n)$ means rescaling  the range  $R(n)$ with the standard deviation $S(n)$. The R/S analysis was popularized by Mandelbrot with his great work on the theory of fractal \cite{Mandelbrot1968noah,Mandelbrot1969robustness,Mandelbrot1982Book} and it was the first kind of time-domain method.
An important result obtained by Mandelbrot is the relation of fractal dimension $D$ and the Hurst exponent $H$, which can be expressed by
\begin{equation} \label{eq-H-D}
D+ H = 2.
\end{equation} 
Equation \eqref{eq-H-D} is important for the Higuchi algorithm for estimating the Hurst index $H$.

\subsection{Integers and Division}

For the positive integers $a, b\in \mathbb{N}$ and $b\ge 2$, we can find an non-negative integer $q\in \mathbb{Z}^+$ such that 
\begin{equation} \label{eq-int-akm}
a = bq+ r, \quad 0 \le r \le b-1
\end{equation}
If $r=0$, then $b$ divides $a$ and $b$ is a factor of $a$,  which can be denoted by $b \mid a$. If $r\neq 0$, then we denote it as
$b\not \mid a$. For $a\ge 2$, if $b\mid a$ implies that $b$ must be $1$ or $a$, then $a$ is called a \textit{prime number}, otherwise it is called a \textit{composite number}.
The factor $b$ of the composite number $a$ such that $b\not\in \set{1,a}$ is called a \textit{proper factor}. 

The integer $q$ in \eqref{eq-int-akm} can be calculated by 
$ q = \mfloor{\frac{a}{b}}$ 
where $\mfloor{x}$ denotes the lower integer of $x$ such that $
\mfloor{x}\le x < \mfloor{x}+1$. 
Similarly, the $\mceil{x}$ denotes the upper integer of $x$ such that
$x \le \mceil{x} < x+1$. For illustration, we have $\mfloor{\pi} = 3$, $\mfloor{-\pi} = -4$, $\mceil{\pi}=4$ and $\mceil{-\pi} = -3$. 

The set of proper factors of the composite number $a\in\mathbb{N}$ can be denoted by
\begin{equation*}
\scrd{S}{pf}(a) = \set{d\in \mathbb{N}: d\mid a, d\ge 2, d \neq a}.
\end{equation*}
Let $a$ be composite number and  $w\in \set{2, \cdots, \mfloor{\sqrt{a}}}$ be a integer for lower bound, then 
\begin{equation}\label{eq-bpf}
\scrd{S}{bpf}(a, w) = \set{d\in \scrd{S}{pf}(a): w\le d\le \mfloor{\frac{a}{w}} }
\end{equation}
is called the set of \textit{bounded proper factors} of $a$. The cardinality of the set $\scrd{S}{bpf}(a, w)$ is denoted by $\card{\scrd{S}{bpf}(a, w)}$, which means the number of the elements in the set. 
For example, we have $\mfloor{\sqrt{48}} = 6$ and $\scrd{S}{pf}(48)=\set{2, 3, 4, 6, 8, 12, 24}$. Thus for $w\in \set{2,3,4, 5, 6}$, equation \eqref{eq-bpf} implies that
\begin{equation*}
\left\{
\begin{array}{ll}
\scrd{S}{bpf}(48,2)=\set{2,3,4,6,8,12, 16, 24}, &\card{\scrd{S}{bpf}(48,2)}=8;\\
\scrd{S}{bpf}(48,3)=\set{3,4,6,8, 12, 16},  &\card{\scrd{S}{bpf}(48,3)}=6;\\
\scrd{S}{bpf}(48,4)=\set{4, 6, 8, 12},   &\card{\scrd{S}{bpf}(48,4)}=4;\\
\scrd{S}{bpf}(48,5)=\set{6, 8},   &\card{\scrd{S}{bpf}(48,5)}=2;\\
\scrd{S}{bpf}(48,6)=\set{6, 8},   &\card{\scrd{S}{bpf}(48,6)}=2.
\end{array}
\right.
\end{equation*}

\subsection{Discrete Fourier Transform}

For the discrete TS $\set{X_1, \cdots, X_{N}}$ with length $N$, its DFT is defined by \cite{Oppenheim3rdDSP}
\begin{equation}\label{eq-DFT}
\hat{X}_k = \sum^{N}_{t=1} X_t \me^{-\frac{2\pi\mi}{N} (k-1)(t-1) }, \quad 1\le k \le N
\end{equation}
where $\mi = \sqrt{-1}$. The sequence $\set{\hat{X}_k: 1\le k\le N}$ is called the frequency spectrum of $\set{X_t: 1\le t\le N}$. The \textit{inverse discrete Fourier transform} (IDFT) for \eqref{eq-DFT} is defined by
 \begin{equation}\label{eq-IDFT}
X_t = \frac{1}{N}\sum^{N}_{k=1} \hat{X}_t \me^{+\frac{2\pi\mi}{N} (k-1)(t-1) }, \quad 1\le t \le N
\end{equation}
Usually the DFT is implemented with the \textit{fast Fourier transform} (FFT)  proposed by Cooley and Tukey in 1965 \cite{FFT1965} in order to reduce the computational complexity from $\mathcal{O}(N^2)$ to $\mathcal{O}(N\log_2 N)$.
The equation \eqref{eq-IDFT} is significant for the spectrum methods for estimating the Hurst exponent and generating the FGN.

\subsection{Discrete Wavelet Transform}

A \textit{discrete wavelet transform} (DWT) is a wavelet transform that decomposes the host signal into wavelets that are discontinuous. Its temporal resolution makes it more attractive over DFT since it contains more information carried both in time and frequency \cite{BEGUM20225856}. 

For the length $N$ of the TS $\set{X_t: 1\le t\le N}$, we can set
the scaling parameter
\begin{equation} \label{eq-Ia}
a \in  \set{1, 2^1, 2^2, \cdots, 2^{J-1}}, \quad J =\mceil{\log_2 N}
\end{equation}
and position parameter
\begin{equation} \label{eq-Ib}
b \in I_a = \set{0, 1, 2, \cdots, a-1 }
\end{equation}
respectively. With the help of \eqref{eq-Ia} and \eqref{eq-Ib}, the scaling functions $ \varphi_{a,b}(t) $ and wavelet functions $ \psi_{a,b}(t) $ are defined by 
\begin{equation}\label{DWT-3}
\left\{
\begin{aligned}
\varphi_{a,b}(t) & =\sqrt{a}\cdot\varphi(a(t-1)-b),\\
\psi_{a,b}(t) &=\sqrt{a}\cdot\psi(a(t-1)-b),
\end{aligned}
\right. 
\quad  1\le t \le N
\end{equation}
Then the DWT of the discrete time signal $ \set{X_t: 1\le t \le N} $ is defined with the scale coefficients $ W_{\vec{X}}^{\varphi}(a,b) $ and detail coefficients $ W_{\vec{X}}^{\psi}(a,b) $, viz. \cite{KEHTARNAVAZ2008175}:
\begin{equation}\label{DWT-1}
\left\{
\begin{aligned}
	W_{\vec{X}}^{\varphi}(a,b)&=\dfrac{1}{\sqrt{N}}\sum_{t=1}^{N}\varphi_{a,b}(t)X_t, \\
	W_{\vec{X}}^{\psi}(a,b)&=\dfrac{1}{\sqrt{N}}\sum_{t=1}^{N}\psi_{a,b}(t)X_t.
\end{aligned}
\right.
\end{equation}
Once the $\varphi_{a,b}(t)$ and $\psi_{a,b}(t)$ are computed by \eqref{DWT-3}, the 
$W_{\vec{X}}^{\varphi}(a,b)$ and $W_{\vec{X}}^{\psi}(a,b)$ can be obtained according to \eqref{DWT-1}. In this paper, only the detail coefficients of the DWT is concerned and we set
\begin{equation}\label{eq-DWT}
	\textrm{DWT}_{b}^{a}(\vec{X}, \psi)=W_{\vec{X}}^{\psi}(a,b).
\end{equation}
Note that the choice of the wavelet function $\psi(t)$ in \eqref{eq-DWT} is not unique for the continuous and discrete wavelet transforms. Both the Harr wavelet \cite{haar1909theorie} and Daubechies wavelet \cite{daubechies1992ten} are satisfactory for estimating the Hurst exponent of the time sequences.

\subsection{Fractional Gaussian Noise}

The \textit{fractional Gaussian noise} (FGN), which is closely associated with the FBM  \cite{Delignieres2011}, is a special type of time sequences with the self-similarity characterized by its autocorrelation. The autocorrelation function of the FGN sequence $U = \set{u_n: n = 0, 1, 2, \cdots}$ is    \cite{Mandelbrot1968fractional}
\begin{equation}\label{fgn-1}
\begin{aligned}
\phi_U(\tau)&=\Expt{u_n u_{n+\tau}}\\
&=\dfrac{1}{2}(|\tau+1|^{2H}\!-2|\tau|^{2H}\!+|\tau-1|^{2H})\\
&=\phi_U(-\tau)
\end{aligned}
\end{equation}  
%\begin{equation}\label{fgn-1}
%\phi_U(\tau)=\Expt{u_n u_{n+\tau}}
%=\dfrac{1}{2}(|\tau+1|^{2H}\!-2|\tau|^{2H}\!+|\tau-1|^{2H})
%=\phi_U(-\tau)
%\end{equation}  
where  $\tau\in \mathbb{Z}^+$ is the time lag.
 With the help of Newton's binomial theorem, we can deduce that 
\begin{equation}\label{fgn-2}
(1\pm\inv{\tau})^{2H} 
= 1 \pm \binom{2H}{1}\cdot \inv{\tau} + H(2H-1)(\pm\inv{\tau})^2 + \mathcal{O}(\tau^{-3})
\end{equation}
for sufficiently large $\tau$. Substitute \eqref{fgn-2} into \eqref{fgn-1}, the autocorrelation function can be expressed by
\begin{equation}\label{fgn-3}
\begin{aligned}
\phi_U(\tau)
&=\dfrac{\tau^{2H}}{2}\left[ (1+\inv{\tau})^{2H}-2+(1-\inv{\tau})^{2H}\right] \\
&=H(2H-1)\tau^{2H-2} + \mathcal{O}(\tau^{-3}).
\end{aligned}
\end{equation}
Equation \eqref{fgn-3} shows that we have the asymptotic property
\begin{equation}\label{fgn-4}
\phi_U(\tau)\propto \tau^{2H-2} \Longleftrightarrow \ln \phi_U(\tau)\sim \scrd{\alpha}{fgn} + \scrd{\beta}{fgn}\ln \tau 
\end{equation}
for the sufficiently large time lag $\tau$ and for the constants $\scrd{\alpha}{fgn}$ and $\scrd{\beta}{fgn} = 2H-2$. 

Theoretically, the FGN sequence is of self-similarity strictly. The property specified by \eqref{fgn-4} can be utilized to generate the FGN sequences with given Hurst exponent \cite{Davies1987tests}. The procedure \ProcName{GenTimeSeqFGN} described in \Algr \ref{alg-fgn} is of great significance for the verification and validation of the estimation algorithms for the Hurst exponent.

\begin{breakablealgorithm}
	\caption{Generating the FGN sequence as the ideal time sequence with specified $ H $.}\label{alg-fgn}
	\begin{algorithmic}[1]
		\Require Sequence length $ \ell $, hurst exponent $ H $.
		\Ensure The FGN sequence $ U = \set{u_t: 0\le t\le \ell} $.
		\Function{GenTimeSeqFGN}{$ \ell, H $}
		%\State $ \vec{k}\gets [0,1,\cdots,\ell-1]$; //  ?????????
		\State $\vec{\rho}\gets \vec{0}\in \ES{R}{1}{\ell}$; // For the autocorrelation function
		\For{$k\in \seq{0, 1, \cdots, \ell-1}$} 
		\State $ \rho_{k+1}\gets0.5\cdot(|{k-1}|^{2H}-2|k|^{2H}+|{k+1}|^{2H})$; 
		\EndFor
		\State $ \vec{g}\gets\FFT\left([\vec{\rho}(1:\ell), 0, \vec{\rho}(\ell:2)]\right) $; // FFT for \eqref{eq-DFT}
		\State $ \vec{V}=\sqrt{\vec{g}} $; // Eigenvalues of the correlation sequence
		\State $\vec{m} \gets \vec{0}\in \ES{R}{1}{\ell}, \vec{n} \gets \vec{0}\in \ES{R}{1}{\ell}$;
		\State $\mu \gets 0, \sigma \gets 1$; \quad // for $\mathcal{N}(0,1)$
		\State $\vec{m} \gets \ProcName{GenTimeSeqGauss}( \mu, \sigma, \ell)$;
		\State $\vec{n} \gets \ProcName{GenTimeSeqGauss}( \mu, \sigma, \ell)$;
		\State $\vec{w}\gets \vec{0} \in \ES{R}{1}{2\ell}$;		
		\State $ w_1 \gets\frac{V_1}{\sqrt{2\ell}}\cdot m_1 $;
		\For{$ j\in\seq{2,3,\cdots,\ell} $}
		\State $w_j\gets{\frac{V_j}{\sqrt{4\ell}}\cdot (m_j+\mi \cdot n_j)}$; // $ \mi=\sqrt{-1} $
		\State $w_{\ell+j}\gets{\frac{V_{\ell+j}}{\sqrt{4\ell}}\cdot (m_{\ell-j+2}-\mi \cdot n_{\ell-j+2})}$;
		\EndFor
		\State $ w_{\ell+1}\gets\frac{V_{\ell+1}}{\sqrt{2\ell}}\cdot n_1 $;
		\State $ \vec{f}\gets\Re(  \FFT(\vec{w})) $; // taking the real part
		\State $ U\gets \ell^{-H}\vec{f}(1:j)$;
		\State\Return $ U$;
		\EndFunction
	\end{algorithmic}
\end{breakablealgorithm}

Note that for the $d$-dimensional vector  $\vec{v}\in \ES{R}{d}{1}$ or $\vec{v}\in \ES{R}{1}{d}$, the notation $\vec{v}(i:r)$ means taking the sub-vector of $\vec{v}$, namely
\begin{equation} \label{eq-vec-sub}
\vec{v}(i:r) = 
\left\{
\begin{aligned}
(v_i, v_{i+1}, \cdots, v_{r-1}, v_{r}), \quad i < r;\\
(v_i, v_{i-1}, \cdots, v_{r+1}, v_{r}), \quad i > r.
\end{aligned}
\right.
\end{equation}
Equation \eqref{eq-vec-sub} is significant in matrix computation and it will be encountered for many times in this paper. 
Furthermore, the procedure $\ProcName{GenTimeSeqGauss}(\mu, \sigma, \ell)$ is used to generate the time sequence with normal distribution in which $\mu$ is the expectation, $\sigma$ is the standard deviation and $\ell$ is the length of the sequence.
Note that the sequences $\vec{m}$ and $\vec{n}$ should be generated independently.

\subsection{Linear Regression and Parameters Estimation}\label{sec-curvefitting}

The estimation of  Hurst exponent is built on the method of linear regression for parameter estimation. Suppose the asymptotic behavior 
of data set $\set{(x_i, y_i): 1\le i \le n}$ can be  expressed by the power law
\begin{equation} \label{eq-power-law}
y_i \propto x_i^\beta \Longleftrightarrow \ln y_i \sim  \alpha + \beta \cdot \ln x_i, \quad \alpha,\beta\in \mathrm{R}
\end{equation}
Let 
\begin{equation}\label{mat-vec-A-b}
	\mat{A}=\begin{bmatrix}
	1&\ln(x_1)\\
	1&\ln(x_2)\\
	\vdots&\vdots\\
	1&\ln(x_n)\\
	\end{bmatrix},\quad \vec{p}=\begin{bmatrix}
	\alpha\\
	\beta\\
	\end{bmatrix},\quad \vec{b}=\begin{bmatrix}
	\ln(y_1)\\
	\ln(y_2)\\
	\vdots\\
	\ln(y_n)
	\end{bmatrix},
\end{equation}
then the data set $\set{(x_i, y_i)}$ specified by \eqref{eq-power-law} can be expressed equivalently with the matrix-vector $\mpair{\mat{A}}{\vec{b}}$ defined by \eqref{mat-vec-A-b}.  
The procedure \ProcName{FormatPowLawData} listed in \Algr \ref{alg-format-PLD} is used for converting the primitive data set with power law  to the  matrix-vector pair $\mpair{\mat{A}}{\vec{b}}$ with standard format.

\begin{breakablealgorithm}
\caption{Converting  primitive data set with power law  to matrix-vector pair} \label{alg-format-PLD} 
\begin{algorithmic}[1] 
\Require Vectors $\vec{x},\vec{y}\in \ES{R}{n}{1}$ such that $y_i\propto x_i^\beta$, sample capacity $n\in \mathbb{N}$ for the data set.
\Ensure Pair $\mpair{\mat{A}}{\vec{b}}$ such that $\mat{A}=(a_{ij})_{2\times n} \in \ES{R}{2}{n}, \vec{b}=(b_i)_{n\times 1}\in \ES{R}{n}{1}$;
\Function{FormatPowLawData}{$\vec{x}, \vec{y}, n$}
\State $\mat{A}\gets \mat{0}_{2\times n}$;
\State $\vec{b}\gets \vec{0}_{n\times 1}$;
\For{$i\in\seq{1, 2, \cdots, n}$}
\State $a_{i1}  \gets 1$;
\State $a_{i2}  \gets \ln(x_i)$;
\State $b_i\gets \ln(y_i)$;
\EndFor
\State \Return $\mpair{\mat{A}}{\vec{b}}$;
\EndFunction 
\end{algorithmic}
\end{breakablealgorithm}

When the pair $\mpair{\mat{A}}{\vec{b}}$ is created, we can establish the following standard over-determined linear system 
\begin{equation} \label{eq-Ax-b-2dim}
\mat{A}\vec{p}=\vec{b}.
\end{equation}
Thus, the parameter vector $\vec{p} = \trsp{[\alpha,\beta]}$ specified by the  \eqref{eq-Ax-b-2dim} can be estimated 
by solving the following convex optimization problem
\begin{equation}\label{eq-popt}
\scrd{\vec{p}}{opt} = \arg \min_{\vec{p}\in \ES{R}{2}{1}} \normp{\mat{A}\vec{p}-\vec{b}}{r},\quad r\in\mathbb{N}
\end{equation}
where $\displaystyle \normp{\vec{x}}{r} =  \sqrt[r]{\abs{x_1}^r + \abs{x_2}^r + \cdots + \abs{x_m}^r}$  denotes the Euclidean norm of the $m$-dim vector $\vec{x} \in\ES{R}{m}{1}$. For $r= 2$ in \eqref{eq-popt}, we can take the
 \textit{least squares} (LS) approach to  solve for $\scrd{\vec{p}}{opt}$ by
\begin{equation}\label{fit-4}
	\scrd{\vec{p}}{LS}=\mat{A}^\dagger\vec{b}
\end{equation}
where the $\mat{A}^\dagger$ in \eqref{fit-4} denotes the Moore-Penrose inverse of the  matrix $\mat{A}$    \cite{ZhangXD2017Matrix}. Various least squares methods can be attempted to obtain the minimum $\ell_2$-norm solution, such as \textit{data least squares} (DLS),  \textit{total least squares} (TLS), and \textit{scaled total least squares} (STLS)    \cite{Paige2002scaled,Zhang2009novel,Zhang2022multi}. For better estimation property, the $\ell_1$-norm for the cost could be used in optimization with the help of residual vector \cite{Fzq2022}.

In the sense of computer programming, the linear regression method of solving the parameter vector $\vec{p}$ in \eqref{eq-Ax-b-2dim} can be encapsulated into a procedure named by \ProcName{LinearRegrSolver} in order to be reused for different applications. The interface can be expressed by
\begin{equation}\label{eq-linsolver}
\vec{p} \gets \ProcName{LinearRegrSolver}(\mat{A}, \vec{b}, n, \cpvar{flag})
\end{equation}
where the $n$ in \eqref{eq-linsolver} is number of data and the \cpvar{flag} is used for selecting the method for solving linear regression problem. For example, $\cpvar{flag} = 2$ implies the $\ell_2$-norm optimization and $\cpvar{flag} = 1$  implies the $\ell_1$-norm optimization.

\subsection{Algorithm for Solving Fixed-Point in Euclidean Space}

For the fixed-point equation 
\begin{equation*} \label{eq-fixedpoint}
\vec{x} = T(f, \vec{x}, \lambda_1, \cdots, \lambda_r), \quad \vec{x}\in \ES{R}{m}{1}
\end{equation*}
where $T$ is a contractive mapping used as the \textit{updating function} in programming language,  $\lambda_1, \cdots, \lambda_r$ are possible extra parameters.  The fixed-point can be solved with an iterative scheme \cite{ZhangHY2024Kuiper}
\begin{equation*}
\vec{x}_{i+1} = T(f, \vec{x}_i, \lambda_1, \cdots, \lambda_r), \quad i = 0, 1, 2, \cdots
\end{equation*}
when the initial value $\vec{x}_0$ and 
 the distance $d(\vec{x}_{i+1}, \vec{x}_i)$, such as the Euclidean norm, are provided properly according to the Cauchy's criteria for convergence.

The pseudo-code for the fixed-point algorithm is listed in  \Algr \ref{alg-fixed-point}, in which 
the concepts of high order function and function object are utilized for the abstraction and flexibility. Note that the order of arguments can be configured by programmers. 
\begin{breakablealgorithm}
\caption{Unified Framework for Solving the Fixed-Point of $\vec{x} = T(f, \vec{x}, \lambda_1, \cdots, \lambda_r)$}\label{alg-fixed-point}
\begin{algorithmic}[1]
\Require Contractive mapping  $T$ as the updator which is a high order function, function object $f$, function object $d$ for the distance $d(\vec{x}_i, \vec{x}_{i+1})$, precision $\epsilon$, initial value $\scrd{\vec{x}}{guess}$ and possible extra parameters $\lambda_1, \cdots, \lambda_r$ with the same or different data types.
\Ensure Fixed-point $\vec{x}$ such that $\vec{x} = T(f, \vec{x}, \lambda_1, \cdots, \lambda_r)$
\Function{FixedPointSolver}{$T, f, d, \epsilon, \scrd{\vec{x}}{guess},  \lambda_1, \cdots, \lambda_r$}
\State $\scrd{\vec{x}}{improve}\gets T(f, \scrd{\vec{x}}{guess}, \lambda_1, \cdots, \lambda_r)$;
\While{$d(\scrd{\vec{x}}{improve}, \scrd{\vec{x}}{guess}) \ge \epsilon$}
\State $\scrd{\vec{x}}{guess} \gets \scrd{\vec{x}}{improve}$;
\State $\scrd{\vec{x}}{improve}\gets T(f, \scrd{\vec{x}}{guess}, \lambda_1, \cdots, \lambda_r)$;
\EndWhile
\State \Return $\scrd{\vec{x}}{improve}$;
\EndFunction
\end{algorithmic}
\end{breakablealgorithm}
In the sense of programming language and discrete mathematics, $f$ is an ordinary (first order) function, $T$ is a second order function and $\ProcName{FixedPointSolve}$ is a third order function. 

The procedure \ProcName{EuclidDist} described by \Algr \ref{alg-calc-dist} is designed for calculating the 
Euclidean distance. For our problem, we have $d(x_i,x_{i+1}) = \abs{x_i-x_{i+1}}$ since it is a 1-dim distance.
\begin{breakablealgorithm}
\caption{Calculate the Euclidean distance of $\vec{x}$ and $\vec{y}$}
\label{alg-calc-dist}
\begin{algorithmic}[1]
\Require $\vec{x}, \vec{y}\in \ES{R}{m}{1}$
\Ensure The Euclidean distance of $\vec{x}$ and $\vec{y}$, i.e., $\displaystyle d(\vec{x},\vec{y})=\normp{\vec{x}-\vec{y}}{2}=\sqrt{\sum^{m}_{i=1}\abs{x_i - y_i}^2}$. 
\Function{EuclidDist}{$\vec{x},\vec{y}$}
\State $\cpvar{sum}\gets 0$;
\For{$\gets i\in \seq{1, \cdots, m}$}
\State $\cpvar{sum}\gets \cpvar{sum} + \abs{x_i-y_i}^2$;
\EndFor
\State $\cpvar{dist} \gets \sqrt{\cpvar{sum}}$; // it will be $d(x,y) = \abs{x-y}$ if $m=1$;
\State \Return $\cpvar{dist}$;
\EndFunction
\end{algorithmic}
\end{breakablealgorithm}

\subsection{Algorithm for Searching a Local Minimum of Single-variable Function}

In 2013, Brent \cite{Brent2013algorithms} provides a line-search method which  is a combination of  golden  section  search  and successive parabolic interpolation . It can be used for finding a local minimum of a single-variable function, please see the \Algr \ref{alg-localmin} in Appendix  \ref{sec-appendix} for details.

\section{Optimal Sequence Partition } \label{sec-opt-seq-partition}

\subsection{Fundamental Operations on Time Sequences}

\subsubsection{Cumulative Sum of Sequence}

For a sequence $\set{x_i: 1\le i \le n}$, its cumulative sequence $\set{c_i: 1\le i \le n}$ is defined by the action of cumulative sum operator on the original sequence. Formally, we have 
\begin{equation} \label{eq-cum-sum-ci}
c_i = \OpCsum{j}{1}{i}{x_j} = \sum^i_{j=1} x_i, \quad 1\le i\le n.
\end{equation}
In consequence, the relation between the cumulative sum operator and arithmetic average operator is
\begin{equation} \label{eq-ave-xi}
\Ave{j}{1}{i}{x_j}  = \frac{1}{i}\cdot\OpCsum{j}{1}{i}{x_j}, \quad 1\le i \le n.
\end{equation} 
Logically, the cumulative sum is more fundamental than the arithmetic average.
Note that \eqref{eq-cum-sum-ci} and \eqref{eq-ave-xi} can be deduced from \eqref{average-operator} directly.

Particularly, for the time sequence $\set{X_t: 1\le t\le N}$, its cumulative sequence is
\begin{equation} \label{eq-ts-cum-sum}
\OpCsum{j}{1}{i}{X_j} = \sum^i_{j=1}X_j, \quad 1\le i \le n
\end{equation}
and its cumulative bias sequence is
\begin{equation} \label{eq-cum-bias-seq}
\OpCsum{j}{1}{i}{X_j-\mean{X}} = \sum^i_{j=1}(X_j-\mean{X}), 
\quad 1\le i \le n
\end{equation}
where the $\mean{X}=\Ave{t}{1}{n}{X_t}$ in \eqref{eq-cum-bias-seq} is the global arithmetic average as described in \eqref{eq-Xt-mean}. The equations \eqref{eq-ts-cum-sum} and \eqref{eq-cum-bias-seq} will be encountered frequently when constructing the estimator of Hurst exponent.

\subsubsection{Sequence Partition}

For computing the Hurst exponent, it is valuable to split the original sequence into several subsequences and construct the statistics of interest.

A fundamental task is to find a suitable length for the subsequence, which  helps in efficiently splitting the sequence as well as preserving the intrinsic characteristics. 

A time sequence $ \vec{X}=\set{X_t}_{t=1}^N $ with length $N$ can be partitioned into $k$ subsequences or segments of equal size $m$ such that
\begin{equation*}
	N = km + r, \quad k = \mfloor{N/m}.
\end{equation*}
If $m\not\mid N$ or equivalently $r\neq 0$, we just ignore the following subsequence with length $r\in \set{1, 2, \cdots, m-1}$, viz.
$$
\set{X_{km+1}, X_{km+2}, \cdots, X_{km+r}}, \quad 1\le r\le  m-1.
$$
The $k$ subsequences of size $m$ obtained can be expressed by
\begin{equation} \label{eq-seq-partition}
	\bigcup^k_{\tau=1}X_{(\tau)} =\set{X_{(\tau)}: 1\le \tau\le k}
\end{equation}
where 
\begin{equation} \label{eq-X-tau-subseq}
	\begin{aligned}
	X_{(\tau)} &= \set{X_{(\tau-1)m+j}: 1\le j \le m}, \quad 1\le \tau \le k\\
	&=\set{X_{(\tau-1)m+1}, \cdots, X_{(\tau-1)m+j}, \cdots, X_{(\tau-1)m+m}}\\
	% &=\vec{X}((i-1)m+1:m)
	\end{aligned}
\end{equation}
is the $\tau$-th subsequence. For simplicity, we can denote the partition operation of the sequence $\vec{X}$ as
\begin{equation} \label{eq-seq-partition-2}
	\set{X_{(\tau)}: 1\le \tau \le k} = \ProcName{SeqPartition}(\vec{X}, m,k)
\end{equation}
according to \eqref{eq-X-tau-subseq}. Equation \eqref{eq-seq-partition-2} can be used to simplify the software implementation of sequence partition.

\subsection{Steps and Algorithms for Optimal Sequence Partition}

\subsubsection{Steps for Optimal Sequence Partition}
According to \eqref{eq-seq-partition}, the partition of $\set{X_t: 1\le N}$ depends on the size $m$ for the subsequences. It is a key issue that how to specify the size $m$. There are three steps for determining the positive integer $m$:
\begin{itemize}
\item generating the set of bounded proper factors for the candidate length and  calculating the cardinality of the set;
\item searching an optimal length for replacing the original sequence length;
\item set the $m$ in the bounded proper factors of the optimal length.
\end{itemize} 

\subsubsection{Brute-force Searching Method for Specifying Bounded Proper Factors of Composite Integer}

The procedure \ProcName{GenSbpf} listed in
\Algr \ref{alg-gen-set-bpf} is designed for finding the set of bounded proper factors of a composite $a$ specified by the integer $w\in\set{2, \cdots, \mfloor{\sqrt{a}}}$.
\begin{breakablealgorithm}
	\caption{Generate the set of bounded proper factors of the composite $a\in \mathbb{N}$ with the lower bound $w$ and the upper bound $a/w$ such that $w\in \set{2, \cdots, \mfloor{\sqrt{a}}}$ with the	brute-force searching method.}\label{alg-gen-set-bpf}
	\begin{algorithmic}[1]
		\Require Composite number $ a \in \mathbb{N}$, lower bound 
		$w\in \set{2, 3, \cdots, \mfloor{\sqrt{a}}}$.
		\Ensure The set $\scrd{S}{bpf}(a, w)$.
		\Function{GenSbpf}{$a, w $}
		\State $ \scrd{S}{bpf}\gets \emptyset $;
		\For{$ i\in\seq{w, w+1, \cdots, \mfloor{a/w}} $}
		\If{ $i\mid a$}
		\State $\scrd{S}{bpf} \gets \scrd{S}{bpf} \cup \set{i}  $;
		\EndIf
		\EndFor
		\State\Return $ \scrd{S}{bpf} $;
		\EndFunction
	\end{algorithmic}
\end{breakablealgorithm}

\subsubsection{Searching Optimal Approximate Length of Sequence}

The procedure \ProcName{SearchOptSeqLen} for searching the optimal length for the subsequence is listed in 
\Algr \ref{alg-opt-seqlen}.

\begin{breakablealgorithm}
	\caption{Searching the optimal length of sequence.}\label{alg-opt-seqlen}
	\begin{algorithmic}[1]
		\Require Sequence length $N$, lower bound $ w\in \set{2, \cdots, \mfloor{\sqrt{N}}}$, percentage $\alpha \in[0.95, 1]$ with default value $\alpha = 0.99$.
		\Ensure Optimal sequence length $\scrd{N}{opt}$ such that $\scrd{N}{opt}\le N$ and $\scrd{S}{pf}(\scrd{N}{opt})$ is not empty.
		\Function{SearchOptSeqLen}{$N, w, \alpha $}
		\State $ \scrd{L}{factors}\gets \emptyset $; // initialize with empty set
		\State $ n_0 \gets \mceil{\alpha N} $;
		\For{$ i\in\seq{n_0, n_0+1, \cdots, N} $}
		\State $ \scrd{S}{bpf}\gets $\ProcName{GenSbpf}$ (i, w) $;
		\State $ \scrd{L}{factors}\gets \scrd{L}{factors} \cup 
		\card{\scrd{S}{bpf}}$; %//添加因子个数$ \card{\scrd{S}{bpf}}$至$ \scrd{L}{factors} $;
		\EndFor
		\State $\mpair{\scrd{i}{max}}{\scrd{v}{max}}\gets\ProcName{SearchMax}(\scrd{L}{factors})$; %// 寻找$ \scrd{L}{factors} $中最大值$\scrd{v}{max}$对应的位置标号$\scrd{i}{max}$;
		\State $ \scrd{N}{opt}\gets n_0+\scrd{i}{max} -1 $;
		\State\Return $ \scrd{N}{opt} $;
		\EndFunction
	\end{algorithmic}
\end{breakablealgorithm}

\subsubsection{Specifying the Parameters of Subsequence}

The $\scrd{N}{opt}$ obtained from the length $N$ must be a composite number. The set of bounded proper factors of $\scrd{N}{opt}$ with lower bound $w$ gives the possible values for the size $m$ required in sequence partition. In other words, we have
\begin{equation} \label{eq-m-set}
m\in \scrd{S}{bpf}(\scrd{N}{opt}, w).
\end{equation}
Once the size $m$ specified by \eqref{eq-m-set} is obtained, the $k = \scrd{N}{opt}/m$ will be determined simultaneously. In consequence, the sequence partition is specified completely. As an illustration, we  take $N=997, w = 20, \alpha = 0.99$, which implies $\scrd{N}{opt} = 990$ by \Algr \ref{alg-opt-seqlen}. Consequently, 
\begin{equation*}
\scrd{S}{bpf}(\scrd{N}{opt}, w) = \scrd{S}{bpf}(990, 22) = \set{22, 30, 33, 45}.
\end{equation*}  
This implies that there are four candidate values for the pair $\mpair{m}{k}$ of interest, i.e.
\begin{align*}
\mpair{m}{k}\in \set{\mpair{22}{45},\mpair{30}{33}, \mpair{33}{30}, \mpair{45}{22}}. 
\end{align*}

\begin{figure*}[htbp]
	\centering
	\includegraphics[width=\textwidth]{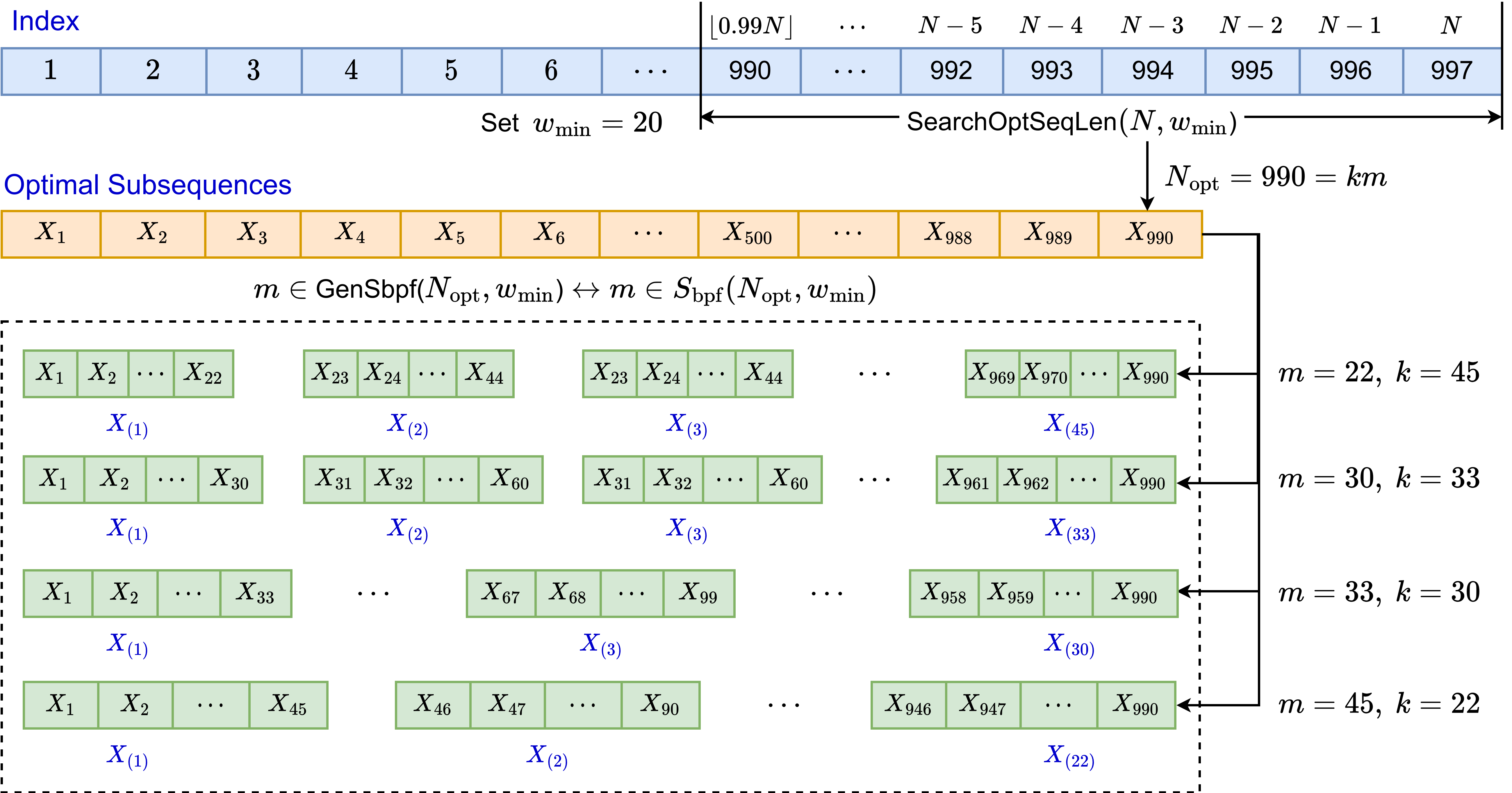}
	\caption{Principle and implementation of optimal sequence partition}
	\label{fig-OptSubSeq}
\end{figure*}

\Fig \ref{fig-OptSubSeq} demonstrates the principle and implementation of optimal sequence partition intuitively. 

\section{Typical Methods For Estimating Hurst Exponent}
\label{sec-typical-methods-for-H}

There are various algorithms for estimating the Hurst exponent $H$ of time sequences based on different principles. For the convenience of 
various applications, it is necessary to describe the algorithms with pseudo-codes and provide the code for the implementation with popular high level programming languages such as C/C++, Python and Octave/MATLAB.
In this section, we will cope with the principles briefly and 
present pseudo-codes for the algorithms for the estimation methods.
We remark that our emphasis is put on the understandability of the principle and the usability of the algorithms for estimating the Hurst exponent of TS.

\subsection{Central Estimation}

\subsubsection{Principle of Central Estimator}

For the time sequence $\vec{X}=\set{X_t: 1\le t\le N}$, we can decompose it into $k$ subsequences with segment size $m$ by \eqref{eq-seq-partition}. Now we compute the moment for each subsequence. Let
\begin{equation} \label{CV-2}
C_\tau = \Ave{j}{1}{m}{X_{(\tau-1)m+j}}, \quad 1\le \tau\le k,
\end{equation}
then the $C_\tau$ in \eqref{CV-2} is the arithmetic average of the subsequence $X_{(\tau)}$. Thus we get 
a new sequence $\set{C_\tau: 1\le \tau \le k}$. The $r$-th central moment of this sequence can be written by
\begin{equation}\label{CV-1}
	\nu(r,m)= \Ave{\tau}{1}{k}{\abs{C_\tau-\mean{X}}^r}.
\end{equation}
where $\mean{X}$ in \eqref{CV-1} is specified by \eqref{eq-Xt-mean}.
If the sequence $ \set{X_t}_{t=1}^N $ is a Gaussian sequence or its variance is finite, then for large $k $ and $m $ the asymptotic property
\begin{equation}\label{CV-3}
	\nu(r,m)\propto m^{r(H-1)}
\end{equation}
holds \cite{Chenjian2006}. Obviously, \eqref{CV-3} means a typical example of power law presented in \eqref{eq-power-law}. By taking the logarithms of both sides of \eqref{CV-3} we immediately have 
\begin{equation} \label{CV-4}
\ln \nu(r,m) \sim \scrd{\alpha}{Central}(r) + \scrd{\beta}{Central}(r) \cdot \ln m
\end{equation}
where
\begin{equation} \label{CV-5}
	\scrd{\beta}{Central}(r)= r(H-1).
\end{equation}
Substituting \eqref{CV-5} into \eqref{CV-4}, we can get the 
 estimator 
\begin{equation} \label{eq-centeralest-order-r}
\scrd{\hat{H}}{Central}(r) = 1 + \frac{1}{r}\scrd{\hat{\beta}}{Central}(r),
\end{equation}
for the Hurst exponent, which can be obtained with the linear regression.
Particularly, for $r=1$ and $r=2$, we have two subtypes of estimation method: 
\begin{itemize}
\item[(1)] \textbf{Absolute Moments} (AM) method \cite{Taqqu1995estimators}, in which $r=1$ and we have 
\begin{equation} \label{eq-H-am}
\scrd{\hat{H}}{AM} = \scrd{\hat{H}}{Central}(1) = 1+ \scrd{\hat{\beta}}{Central}(1).
\end{equation} 
\item[(2)] \textbf{Aggregate Variance} (AV) method \cite{Taqqu1995estimators}, in which  $r=2$ and we have
\begin{equation} \label{eq-H-av}
\scrd{\hat{H}}{AV} = \scrd{\hat{H}}{Central}(2) = 1+ \frac{1}{2}\scrd{\hat{\beta}}{Central}(2).
\end{equation}
\end{itemize}
Obviously, \eqref{eq-H-am} and \eqref{eq-H-av} are special cases of \eqref{eq-centeralest-order-r}.

\subsubsection{Algorithm for Central Estimator}

The key step of central estimation is constructing the corresponding central moments $\nu(r,m)$ according to \eqref{CV-1}. The procedure
\ProcName{EstHurstCentral} listed in \Algr \ref{alg-Central} is used to estimate the Hurst exponent by \eqref{CV-2}, \eqref{CV-1}, \eqref{CV-3}, \eqref{CV-4}, \eqref{CV-5} and \eqref{eq-centeralest-order-r}, which relies on two procedures:
\begin{itemize}
\item the procedures \ProcName{GenSbpf} listed in \Algr \ref{alg-gen-set-bpf} for  generating factor set, and
\item the procedure \ProcName{SearchOptSeqLen} listed in \Algr \ref{alg-opt-seqlen} for searching the optimal sequence length.
\end{itemize} 
For time-domain methods, the parameter estimation for linear model is fundamental for the estimation, please see the subsection  \ref{sec-curvefitting} about linear regression.

\begin{breakablealgorithm}
	\caption{Central Estimator for Hurst exponent}\label{alg-Central}
	\begin{algorithmic}[1]
		\Require Time sequence $ \vec{X} $, window size $ w $, order $ r\in \set{1,2} $, indicator \cpvar{flag} for the optimization method in linear regression.
		\Ensure Hurst exponent of the sequence $ \vec{X} $.
		\Function{EstHurstCentral}{$ \vec{X}, w,r,\cpvar{flag} $}
		\State $ N \gets  \ProcName{GetLength}(\vec{X})$;% // Length of original sequence
		\State $ \scrd{N}{opt}\gets \ProcName{SearchOptSeqLen}(N, w) $;
		\State $ \vec{T}\gets \ProcName{GenSbpf}(\scrd{N}{opt}, w) $; // $\scrd{S}{bpf}(\scrd{N}{opt},w)$
		\State $ n\gets  \ProcName{GetLength}(\vec{T})$; // $\card{\scrd{S}{bpf}(\scrd{N}{opt},w)}$
		\State $ \vec{S}\gets\vec{0}\in\ES{R}{n}{1} $; // For the statistics
		\State $ \mean{X} \gets\Ave{t}{1}{N}{X_t} $;
		\For{$ \cpvar{idx}\in\seq{1,2,\cdots,n} $}
		\State $ m\gets \scrd{T}{idx} $; // $ m $ is the interval time
		\State $ k\gets\mfloor{\scrd{N}{opt}/m} $; // number of subsequences
        \State $\vec{Y}\gets \vec{0}\in \ES{R}{k}{1}$; // $\vec{Y}=\trsp{[Y_1,\cdots,Y_\tau,\cdots,Y_k]}$;
        \State $\set{X_{\tau}:1\le \tau\le k} \gets \ProcName{SeqPartition}(\vec{X},N, m)$;        
        \For{$\tau\in\set{1,2,\cdots,k}$}		
		\State $ {Y}_\tau \gets\Ave{j}{1}{m}{X_{(\tau-1)m+j}}$; // arithmetic ave
		\EndFor
		\If{$ r=1 $}
		\State $ \nu\gets\normp{\vec{Y}-\mean{X}}{1}/k $; \quad // $\ell_1$-norm here
		\Else
		\State $ \nu\gets\Var(\vec{Y}) $; \quad // $r = 2$
		\EndIf
		\State $ \scrd{S}{idx}\gets\nu $;
		\EndFor
		\State $ \seq{\mat{A},\vec{b}}\gets\ProcName{FormatPowLawData}(\vec{T},\vec{S},n) $;
		\State $\vec{p}\gets \ProcName{LinearRegrSolver}(\mat{A}, \vec{b}, n, \cpvar{flag})$; // by \eqref{eq-linsolver}
		\State $\scrd{\beta}{Central} \gets p_2$ ; // $\vec{p} = \trsp{[\alpha,\beta]}$;
		%\State Fitting curve $ \ln (\scrd{S}{m})\sim\ln (\scrd{S}{c}) $ obtained the  slope as $ \scrd{\beta}{Central} $;
		\State $H \gets \scrd{\beta}{Central}/r + 1$;
		\State\Return $ H $;
		\EndFunction
	\end{algorithmic}
\end{breakablealgorithm}

\subsection{Generalized Hurst Exponent Method (GHE)}

\subsubsection{Principle of GHE Estimator}
For time sequence $ \set{X_t}_{t=1}^N $, the $q$-th order moment of the distribution of the increments  based on the time lag $ \tau $ can be written by  \cite{Barabasi1991multifractality}
\begin{equation}\label{GHE-1}
	\mu_q(\tau)=\Ave{i}{1}{N-\tau}{\abs{X_{i+\tau}-X_i}^q}
\end{equation}
The \textit{generalized Hurst exponent} (GHE), denoted by $ H(q)$, can be deduced from the asymptotic scaling behavior of \eqref{GHE-1}, viz.  \cite{Di2003scaling}
\begin{equation}\label{GHE-2}
	\mu_q(\tau)\propto \tau^{qH(q)} \Longleftrightarrow \ln \mu_q(\tau) \sim \scrd{\alpha}{GHE} +
	\scrd{\beta}{GHE} \cdot \ln \tau
\end{equation}
Obviously, \eqref{GHE-2} is a special case of the equivalent form of power law specified by \eqref{eq-power-law}. In consequence, the generalized Hurst exponent $ H(q) $  can be expressed by  \cite{Di2003scaling}
\begin{equation}\label{GHE-4}
\scrd{\hat{H}}{GHE} = \frac{\scrd{\hat{\beta}}{GHE}}{q}. 
\end{equation}

\subsubsection{Algorithm for GHE Estimator}

The procedure \ProcName{EstHurstGHE} described in \Algr \ref{alg-GHE}  can be used to compute the Hurst exponent based on \eqref{GHE-1}, \eqref{GHE-2} and \eqref{GHE-4}.

\begin{breakablealgorithm}
	\caption{Generalized Hurst Exponent Method}\label{alg-GHE}
	\begin{algorithmic}[1]
		\Require Time series data $ \vec{X} $, order $ q $, indicator \cpvar{flag} for the optimization method in linear regression.
		\Ensure Hurst exponent of the sequence $ \vec{X} $.
		\Function{EstHurstGHE}{$ \vec{X}, q, \cpvar{flag}$}
		\State $ n\gets10 $;
		\State $ N \gets  \ProcName{GetLength}(\vec{X}) $;
		\State $ \vec{T}\gets\trsp{[1,2,\cdots,n]}\in \ES{R}{n}{1} $;
		\State $ \vec{S}\gets\vec{0}\in\ES{R}{n}{1} $; // For the statistics
		\State $ \mean{X}\gets\Ave{i}{1}{N}{X_i} $; 
		\State $\vec{Y}\gets \vec{0}\in \ES{R}{N}{1}$;
		\For{$i\in \seq{1, 2, \cdots, N}$}
		\State $ Y_i\gets \OpCsum{j}{1}{i}{X_{j}-\mean{X}}$;// by \eqref{eq-cum-bias-seq}
		\EndFor
		\For{$ \cpvar{idx}\in\seq{1,2,\cdots,n} $}
		\State $ \mu_q\gets\Ave{i}{1}{N-\cpvar{idx}}{\abs{Y_{i+\cpvar{idx}}-Y_i}^q} $;// by \eqref{GHE-1}
		\State $ S_\cpvar{idx}\gets\mu_q $;
		\EndFor
		\State $ \seq{\mat{A},\vec{b}}\gets\ProcName{FormatPowLawData}(\vec{T},\vec{S},n) $;
		\State $\vec{p}\gets \ProcName{LinearRegrSolver}(\mat{A}, \vec{b}, n, \cpvar{flag})$;
		\State $\scrd{\beta}{GHE} \gets p_2$ ; // $\vec{p} = \trsp{[\alpha,\beta]}$;
		\State $ H\gets \scrd{\beta}{GHE}/q $;// by \eqref{GHE-4}
		\State \Return $ H $;
		\EndFunction
	\end{algorithmic}
\end{breakablealgorithm}

\subsection{Higuchi Method (HM)}

\subsubsection{Principle of Higuchi Estimator}

For the time sequence $\set{X_t}_{t=1}^N $, its cumulative bias sequence is defined by   \cite{Taqqu1995estimators}
\begin{equation}\label{HM-1}
	Y_i = \OpCsum{j}{1}{i}{X_j-\mean{X}}, \quad 1\le i\le N
\end{equation}
where $\mean{X} = \Ave{t}{1}{N}{X_t}$ and the normalized length of each sample 
\begin{equation}\label{HM-2}
	L_k(m)=\dfrac{\gamma}{m}\sum\limits^{\mfloor{(N-k)/{m}}}_{i=1}\abs{Y_{k+im}-Y_{k+(i-1)m}}, \quad 1 \le k \le m
\end{equation}
can be calculated with the help of \eqref{HM-1}, where the integers $k$ and $m $ indicate the initial time and the interval time respectively. The normalization factor $\gamma$ in \eqref{HM-2} is defined by
\begin{equation}\label{HM-3}
	\gamma = \dfrac{N-1}{\mfloor{(N-k)/{m}}\cdot m}. 
\end{equation}
Higuchi showed that \cite{Higuchi1988approach} 
\begin{equation}\label{HM-4}
	L(m)\propto m^{-D} \Longleftrightarrow \ln L(m) \sim \scrd{\alpha}{HM} + \scrd{\beta}{HM} \ln m
\end{equation}
where the parameter $D$ is the fractal dimension of the time sequence, $\scrd{\beta}{HM} = -D$ and
 \begin{equation*}
L(m)=\Ave{k}{1}{m}{L_k(m)}.
\end{equation*}
Obviously, \eqref{HM-4} is a special case of the equivalent form of power law specified by \eqref{eq-power-law}.
Substituting the \eqref{eq-H-D} into \eqref{HM-4}, we can obtain the Higuchi estimator 
\begin{equation}\label{HM-6}
\scrd{\hat{H}}{HM} = 2 + \scrd{\hat{\beta}}{HM}.
\end{equation}
for the Hurst exponent according to \eqref{HM-4}.
\subsubsection{Algorithm for Higuchi Estimator}

The procedure \ProcName{EstHurstHiguchi} described \Algr \ref{alg-HM} is designed for estimating the Hurst exponent based on \eqref{HM-1}, \eqref{HM-2}, \eqref{HM-3},\eqref{HM-4} and \eqref{HM-6}. 
\begin{breakablealgorithm}
	\caption{Higuchi method for estimating the Hurst exponent}\label{alg-HM}
	\begin{algorithmic}[1]
		\Require Time series data $ \vec{X} $, indicator \cpvar{flag} for the optimization method in linear regression.
		\Ensure Hurst exponent of the sequence $ \vec{X} $.
		\Function{EstHurstHiguchi}{$ \vec{X}, \cpvar{flag} $}
		\State $ n\gets10 $;
		\State $ N \gets  \ProcName{GetLength}(\vec{X}) $;% // Length of original sequence
		\State $ \vec{T}\gets\vec{0}\in\ES{R}{n}{1} $; // For the interval time
		\State $ \vec{S}\gets\vec{0}\in\ES{R}{n}{1} $; // For the statistics
		\State $ \mean{X}\gets\A(X_i) $; 
		\State $\vec{Y}\gets \vec{0}\in \ES{R}{N}{1}$;
		\For{$i\in \seq{1, 2, \cdots, N}$}
		\State $ Y_i\gets \OpCsum{j}{1}{i}{X_{j}-\mean{X}}$;// by \eqref{HM-1}
		\EndFor
		\For{$ \cpvar{idx}\in\seq{1,2,\dots,n} $}
		\If{$ \cpvar{idx}>4 $} 
		\State $m\gets{\mfloor{2^{{(\cpvar{idx}+5)/4}}}}$;
		\Else
		\State $ m\gets\cpvar{idx} $;
		\EndIf
		\State $T_{\cpvar{idx}}\gets m$;
		\State $ k\gets\mfloor{{N}/{m}}$;
		\State $ {L_k}\gets\Ave{i}{1}{k-1}{\Ave{j}{(i-1)m+1}{im}{\left|Y_{j+m}\!-\!Y_{j}\right|}} $;
		\State $ S_{\cpvar{idx}} \gets (N-1)\cdot L_k/{m^2}$; // by \eqref{HM-2} and \eqref{HM-3}
		\EndFor
		\State $ \seq{\mat{A},\vec{b}}\gets\ProcName{FormatPowLawData}(\vec{T},\vec{S},n) $;
		\State $\vec{p}\gets \ProcName{LinearRegrSolver}(\mat{A}, \vec{b}, n, \cpvar{flag})$;
		\State $\scrd{\beta}{HM} \gets p_2$ ; // $\vec{p} = \trsp{[\alpha,\beta]}$;
		\State $H \gets \scrd{\beta}{HM}+2$; // by \eqref{HM-6}
		\State\Return $ H $;
		\EndFunction
	\end{algorithmic}
\end{breakablealgorithm}

\subsection{Detrended Fluctuation Analysis (DFA)}

\subsubsection{Principle of DFA Estimator}

The DFA method for computing the Hurst exponent is based on the sequence partition. For the time sequence $ \set{X_t}_{t=1}^{N} $, with the configuration of parameter $\alpha\in[0.95,1]$ and minimal size $w$ of the subsequence, then optimal sequence length $\scrd{N}{opt}$ can be solved with \Algr \ref{alg-opt-seqlen}, the size $m$ can be calculated by \eqref{eq-m-set} and the number of subsequence will be $k = \scrd{N}{opt}/m$.

\begin{figure*}[htbp]
\centering
\includegraphics[width=0.9\textwidth]{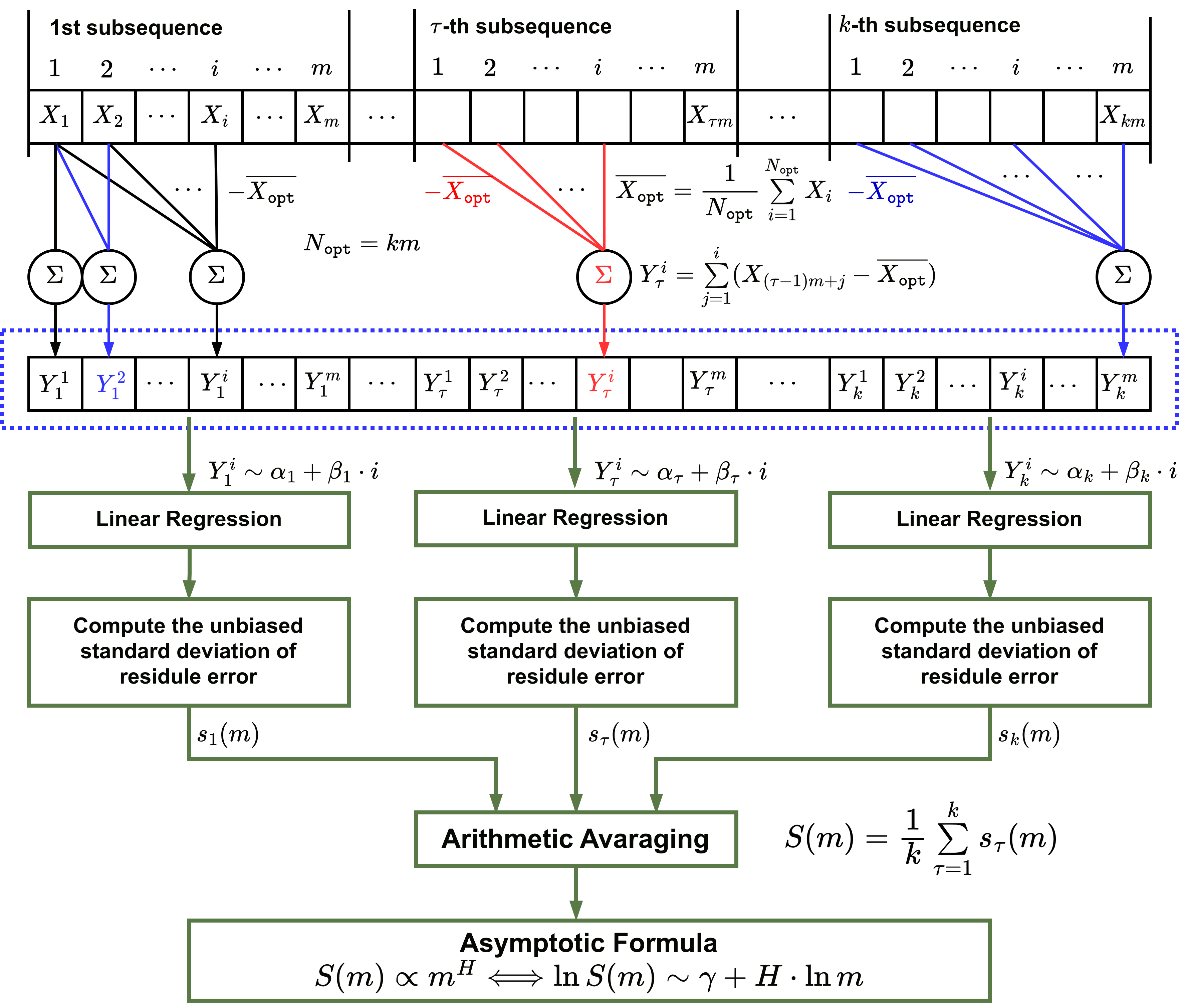} 
\caption{Partition of time sequence and structure of cumulative sequence for DFA method}
\label{fig-DFA}
\end{figure*} 

For the purpose of clarity and intuition, the principle DFA method is shown in \Fig \ref{fig-DFA}.
We now give some interpretations for the steps for the DFA estimator: \begin{itemize}
\item[0)] Pre-conditioning: \\
Partitioning the sequence $\set{X_t: 1\le t\le N}$ into 
$k$ subsequences with minimal size $w$ such that $\scrd{N}{opt} = mk$ and 
$ m\in \scrd{S}{bpf}(\scrd{N}{opt},w)=\set{m_1, m_2, \cdots, m_n}
$  where $n = \card{\scrd{S}{bpf}(\scrd{N}{opt},w)}$.
\item[i)] Computing the global arithmetic average of the optimal sequence $\set{X_t: 1\le t\le \scrd{N}{opt}=mk}$ by
\begin{equation} \label{DFA-1}
\mean{\scrd{X}{opt}} = \Ave{t}{1}{\scrd{N}{opt}}{X_t}
\end{equation}
and construct the cumulative bias sequence 
\begin{equation} \label{DFA-2}
Z_i = \OpCsum{j}{1}{i}{X_j-\mean{\scrd{X}{opt}}}, \quad  1\le i \le \scrd{N}{opt}.
\end{equation}
\item[ii)] Constructing the $k$ subsequences $\set{Y_\tau^i: 1\le i \le m}$
for $1\le \tau \le k$ by 
\begin{equation} \label{DFA-3}
\vec{Y}_\tau = \trsp{[Y_\tau^1, Y_\tau^2, \cdots, Y_\tau^m]}, \quad 1\le \tau\le k
\end{equation}
where
\begin{equation}\label{DFA-4}
Y_\tau^i= Z_{(\tau-1)m+i}, \quad 1\le i\le m. 
\end{equation}
\item[iii)] Performing the linear regression for each subsequence $\set{Y^i_\tau: 1\le i\le m}$ specified by \eqref{DFA-1} for $1\le \tau \le k$  \cite{Peng1994mosaic,Weron2002estimating}
\begin{equation}\label{DFA-5}
	Y^i_\tau\sim \alpha_\tau+\beta_\tau\cdot i, \quad 1\le i\le m
\end{equation}
or equivalently
\begin{equation} \label{DFA-6}
\vec{q}_\tau \gets 
\ProcName{LinearRegrSolver}(\mat{M},\vec{Y}_\tau,m,\cpvar{flag})
\end{equation}
for $\vec{q}_\tau =\trsp{[\alpha_\tau,\beta_\tau]}$ such that
\begin{equation} \label{DFA-7}
\mat{M} = \trsp{\begin{bmatrix}
1 & 1  & \cdots & 1\\
1 & 2 & \cdots & m\\
\end{bmatrix}}, \quad \vec{\varepsilon}_\tau = \mat{Y}_\tau - \mat{M}\vec{q}_\tau
\end{equation}
or equivalent in the component form 
\begin{equation} \label{DFA-7b}
\varepsilon_\tau^i = Y^i_\tau-\alpha_\tau- \beta_\tau\cdot i, \quad 1\le i\le m.
\end{equation}
\item[iv)] Calculating the unbiased standard deviation of the residual sequence
\begin{equation}\label{DFA-8}
s_\tau(m)=\OpStd{i}{1}{m}{\varepsilon_\tau^i}.
%=\sqrt{\dfrac{1}{m-1}\sum_{i=1}^{m}(\varepsilon_\tau^i-\mean{\varepsilon})^2}.
\end{equation}
\item[v)] Calculating the arithmetic average of each standard deviation
\begin{equation}\label{DFA-9}
	S(m)=\Ave{\tau}{1}{k}{s_\tau(m)}.
\end{equation}
in order to get the asymptotic relation
\begin{equation} \label{DFA-10}
 S(m)\propto m^H \Longleftrightarrow \ln S(m) \sim \gamma+ H \cdot \ln m
\end{equation}
\item[vi)] Repeating the steps i) $\sim$ v) for $n$ times for different choices of $m$ and setting
\begin{equation*}
\left\{
\begin{aligned}
&\vec{T}=\trsp{[m_1, \cdots, m_n]}\\
&\vec{S}=\trsp{[S(m_1), \cdots, S(m_n)]}\\
&\mpair{\mat{A}}{\vec{b}}\gets \ProcName{FormatPowLawData}(\vec{T},\vec{S},n)
\end{aligned}
\right.
\end{equation*}
\item[vii)] Estimating the Hurst exponent with linear regression 
\begin{equation*}
\scrd{\vec{p}}{DFA}\gets \ProcName{LinearRegrSolver}(\mat{A},\vec{b},k,\cpvar{flag})
\end{equation*} 
where $\scrd{\vec{p}}{DFA}=\trsp{[\scrd{\hat{\alpha}}{DFA},\scrd{\hat{\beta}}{DFA}]}$, which implies that 
\begin{equation*}
\scrd{\hat{H}}{DFA} = \scrd{\hat{\beta}}{DFA}.
\end{equation*} 
\end{itemize}
\subsubsection{Algorithm for DFA Estimator}

The curve fitting method is a crucial step in the DFA-method, we need to frequently solve for the slope and intercept to construct the corresponding residual vectors. The \Algr \ref{alg-DFA} is designed to compute the Hurst exponent with the DFA estimator based on \eqref{DFA-1}, \eqref{DFA-2}, \eqref{DFA-3}, \eqref{DFA-4}, \eqref{DFA-5}, \eqref{DFA-6}, \eqref{DFA-7}, \eqref{DFA-7b}, \eqref{DFA-8}, \eqref{DFA-9} and \eqref{DFA-10}.

\begin{breakablealgorithm}
	\caption{Detrended Fluctuation Analysis Estimator for Hurst exponent}\label{alg-DFA}
	\begin{algorithmic}[1]
		\Require Time sequence $ \vec{X} $, window size $ w $, indicator \cpvar{flag} for the optimization method in linear regression.
		\Ensure Hurst exponent of the sequence $ \vec{X} $.
		\Function{EstHurstDFA}{$ \vec{X},w, \cpvar{flag} $}
		\State $ N \gets  \ProcName{GetLength}(\vec{X}) $;
		\State $ \scrd{N}{opt}\gets\ProcName{SearchOptSeqLen}(N,w) $;
		\State $\vec{T}\gets \ProcName{GenSbpf}(\scrd{N}{opt}, w)$; // $\scrd{S}{bpf}(\scrd{N}{opt},w)$
		\State $ n\gets \ProcName{GetLength}(\vec{T})$; // $\card{\scrd{S}{bpf}(\scrd{N}{opt},w)}$
		\State $ \vec{S}\gets\vec{0}\in\ES{R}{n}{1} $; // For the statistics
		\State $\vec{Z}\gets \vec{0}\in \ES{R}{N}{1}$; // global cumulative sequence
		\State $ \mean{\scrd{X}{opt}}\gets\Ave{i}{1}{\scrd{N}{opt}}{X_i}$; // global arithmetic average
		\For{$ i \in\seq{1, 2, \cdots, N}$}
		\State $ Z_i\gets \OpCsum{j}{1}{i}{X_j-\mean{\scrd{X}{opt}}}$;
		\EndFor
		\For{$ \cpvar{idx}\in\seq{1,2,\cdots,n} $}
		\State $ m\gets T_{\cpvar{idx}} $;
		\State $ k\gets\scrd{N}{opt}/m $;
		\State $ \vec{s}_\tau\gets\vec{0}\in\ES{R}{k}{1} $; // vector of standard deviation
		\State $ \vec{\varepsilon}_\tau \gets\vec{0}\in\ES{R}{m}{1} $; // vector of regression residuals
		\State $ \mat{M}\gets \trsp{\begin{bmatrix}
			1 & 1  & \cdots & 1\\
			1 & 2 & \cdots & m\\
			\end{bmatrix}}$; // for linear regression
		\State $\vec{q}_\tau\gets \vec{0}\in \ES{R}{2}{1}$; //  $\vec{q}=\trsp{[\alpha,\beta]}$
		\For{$ \tau\in\seq{1,2,\dots,k} $}
		\State $ \vec{Y}_\tau \gets\trsp{[Z_{(\tau-1)m+1}, Z_{(\tau-1)m+2}, \cdots, Z_{\tau m}]} $;
		\State $\vec{q}_\tau\gets \ProcName{LinearRegression}(\mat{M},\vec{Y}_\tau, m, \cpvar{flag})$;
        \State $\vec{\varepsilon}_\tau \gets \vec{Y}_\tau - \mat{M}\vec{q}_\tau $;
		\State $ s_\tau \gets \OpStd{i}{1}{m}{\varepsilon_\tau^i}$; 
		\EndFor
		\State $ S_{\cpvar{idx}} \gets \Ave{\tau}{1}{k}{s_\tau}$;
		\EndFor
		\State $ \seq{\mat{A},\vec{b}}\gets\ProcName{FormatPowLawData}(\vec{T},\vec{S},n) $;
		\State $\vec{p}\gets \ProcName{LinearRegrSolver}(\mat{A}, \vec{b}, n, \cpvar{flag})$;
		\State $\scrd{\beta}{DFA} \gets p_2$ ; // $\vec{p} = \trsp{[\alpha,\beta]}$;
		\State $ H\gets \scrd{\beta}{DFA} $;
		\State \Return $ H $;
		\EndFunction
	\end{algorithmic}
\end{breakablealgorithm}

It should be noted that in \Algr \ref{alg-DFA}, we have taken a strategy of cumulative calculation over the entire sequence. By comparing with the operation of cumulative calculation for each samples as described in equation \eqref{DFA-4}, this approach  can reduce the computational complexity significantly. 

\subsection{Rescaled Range Analysis (R/S Analysis)}

\subsubsection{Principle of R/S Estimator}

\begin{figure*}[htbp]
	\centering
	\includegraphics[width=0.9\textwidth]{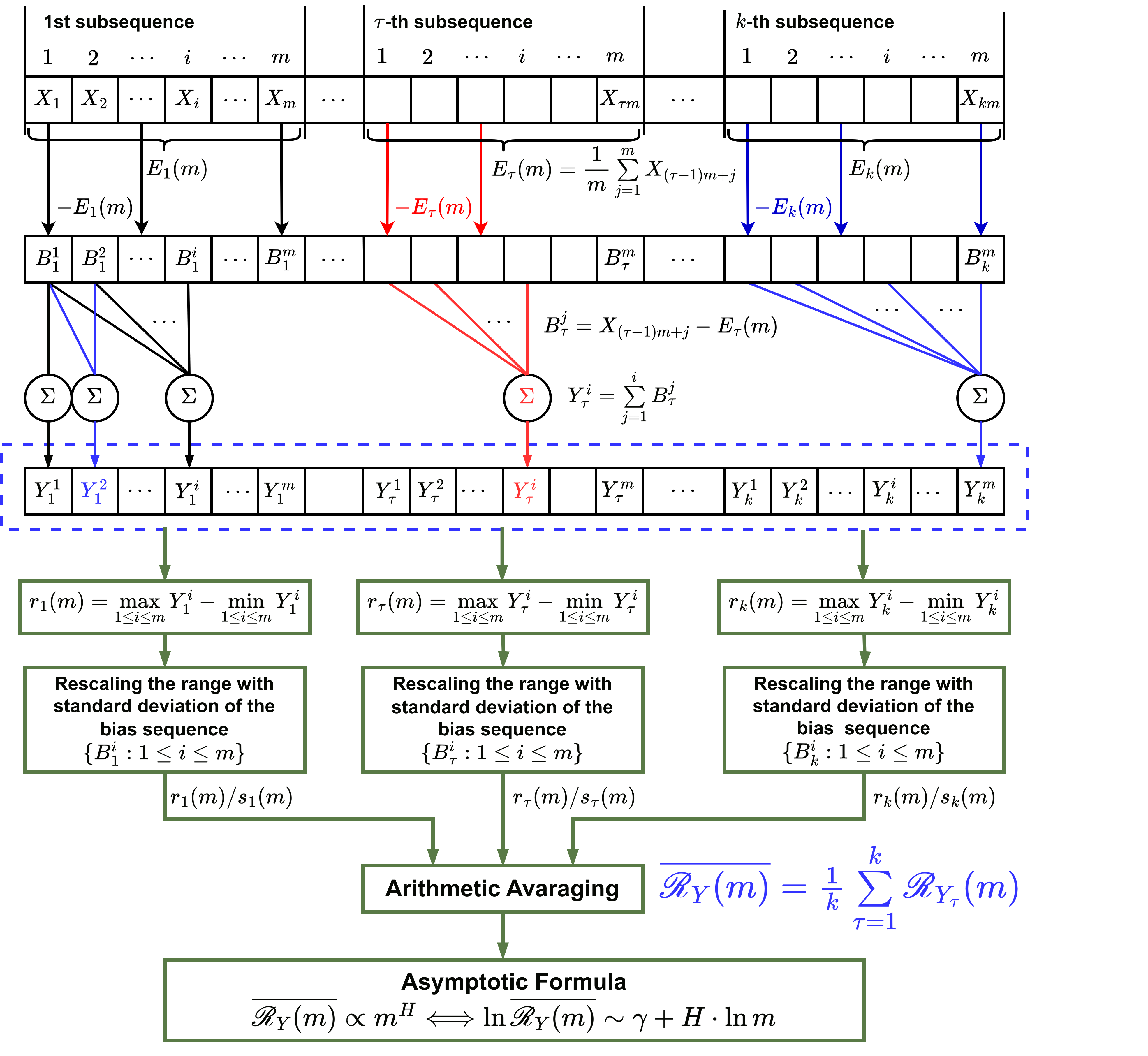} 
	\caption{Partition of time sequence and structure of cumulative sequence for R/S analysis}
	\label{fig-seq-partition-RS}
\end{figure*}

Similar to the DFA method, for the time sequence $ \set{X_t}_{t=1}^{N} $, with the configuration of parameter $\alpha\in[0.95,1]$ and minimal size $w$ of the subsequence, the optimal sequence length $\scrd{N}{opt}$ can be solved with \Algr \ref{alg-opt-seqlen}, the size $m$ can be calculated by \eqref{eq-m-set} and the number of subsequence will be $k = \scrd{N}{opt}/m$. The way for finding the Hurst exponent \cite{Weron2002estimating} by R/S method is illustrated in \Fig \ref{fig-seq-partition-RS}.  

Here we give some interpretations for the steps of the R/S estimator:
\begin{itemize}
\item[0)] Pre-conditioning: Partitioning the sequence $\set{X_t: 1\le t\le N}$ into 
$k$ subsequences with minimal size $w$ such that $\scrd{N}{opt} = mk$ and 
$m\in \scrd{S}{bpf}(\scrd{N}{opt},w)=\set{m_1, m_2, \cdots, m_n}$ 
where $n = \card{\scrd{S}{bpf}(\scrd{N}{opt},w)}$.
\item[i)] For each subsequence,  computing its local arithmetic average  by
\begin{equation}\label{RS-1}
E_\tau(m)=\Ave{j}{1}{m}{X_{(\tau-1)m+j}}.
\end{equation}
\item[ii)] Construct the $\tau$-th local biased/detrended sequence
\begin{equation} \label{RS-2}
B_\tau=\set{B_\tau^j: 1\le j\le m}
\end{equation} 
in which 
\begin{equation} \label{RS-3}
B_\tau^j = X_{m(\tau-1)+j}-E_\tau(m), \quad 1\le j \le m 
\end{equation}
and the cumulative bias sequence
\begin{equation} \label{RS-4}
Y_\tau=\set{Y_\tau^i: 1\le i\le m}
\end{equation}
where
\begin{equation} \label{RS-5}
Y_\tau^i=\OpCsum{j}{1}{i}{B_\tau^j} = \sum\limits_{j=1}^{i} B_\tau^j, \quad 1\le i\le m. 
\end{equation}
\item[iii)] Calculating the unbiased standard deviation of the local bias sequence
\begin{equation}\label{RS-6}
s_\tau(m)= \OpStd{j}{1}{m}{B_\tau^j}, \quad 1\le \tau \le k
\end{equation}
\item[iv)] Calculating the range for the $\tau$-th cumulative bias  sequence $Y_\tau$
\begin{equation}\label{RS-7}
r_\tau(m)=\max_{1\le i\le m}Y_\tau^i- \min_{1\le i\le m}Y_\tau^i, \quad 1\le \tau\le k
\end{equation}
\item[v)] Computing the R/S statistics of the sequence $Y_\tau$
\begin{equation}\label{RS-8}
\RRS{Y_\tau}(m)=\frac{r_\tau(m)}{s_\tau(m)} 
\end{equation} 
\item[vi)] Calculating the arithmetic average of each R/S statistics
\begin{equation}\label{RS-9}
\mean{\RRS{Y}(m)}=\Ave{\tau}{1}{k}{\RRS{Y_\tau}(m)} = \frac{1}{k}\sum^k_{\tau=1} \frac{r_\tau(m)}{s_\tau(m)}. 
\end{equation}
in order to get the asymptotic relation
\begin{equation*}
\mean{\RRS{Y}(m)}\propto m^H\Longleftrightarrow \ln 
\mean{\RRS{Y}(m)} \sim \gamma + H \cdot \ln m
\end{equation*}
\item[vii)] Repeating the steps i)$\sim$ v) for $n$ times for different choices of $m$ and setting
\begin{equation*}
\left\{
\begin{aligned}
&\vec{T}=\trsp{[m_1, \cdots, m_n]}\\
&\vec{S}=\trsp{[S(m_1), \cdots, S(m_n)]}\\
&\mpair{\mat{A}}{\vec{b}}\gets \ProcName{FormatPowLawData}(\vec{T},\vec{S},n)
\end{aligned}
\right.
\end{equation*}
\item[viii)] Estimating the Hurst exponent with linear regression 
\begin{equation*}
\scrd{\vec{p}}{RS}\gets \ProcName{LinearRegrSolver}(\mat{A},\vec{b},k,\cpvar{flag})
\end{equation*} 
where $\scrd{\vec{p}}{RS}=\trsp{[\scrd{\hat{\alpha}}{RS},\scrd{\hat{\beta}}{RS}]}$, which implies that 
\begin{equation*}
\scrd{\hat{H}}{RS} = \scrd{\hat{\beta}}{RS}.
\end{equation*}
\end{itemize}

Particularly, the theoretical values of the R/S statistics of white noise are usually approximated by   \cite{Annis1976expected}
\begin{equation}\label{RS-10}
	\Expt{\RRS{Y}(m)}\!=\left\{\!\!
	\begin{array}{ll}
	\frac{m-\frac{1}{2}}{m} \cdot\frac{\Gamma\left(\frac{m-1}{2}\right)}{\sqrt{\pi}\cdot \Gamma\left(\frac{m}{2}\right)}\cdot\sum\limits_{i=1}^{m-1}\sqrt{\frac{m-i}{i}}, &
	 n\le 340\\
	\frac{m-\frac{1}{2}}{m} \cdot\sqrt{\frac{2}{\pi m}}\cdot\sum\limits_{i=1}^{m-1}\sqrt{\frac{m-i}{i}}, & n>340
	\end{array}
	\right.
\end{equation}
where the $\displaystyle \Gamma(x)=\int^{+\infty}_{0}t^{x-1}\me^{-t}\dif t$ in \eqref{RS-10} is the Gamma function and the factor $ (m-1/2)/m $  was added by Peters   \cite{Peters1994fractal} to improve the accuracy for small $ m $. Thus we can construct the revised R/S statistics  \cite{Weron2002estimating}:
\begin{equation}\label{RS-11}
\mathscr{R}_{Y}^{\mathrm{AL}}(m) = \RRS{Y}(m)-\Expt{\RRS{Y}(m)}+\sqrt{\frac{\pi m}{2}}
\end{equation}
for the $Y_\tau$ which has the asymptotic behavior 
\begin{equation}\label{RS-12}
\mathscr{R}_{Y}^{\mathrm{AL}}(m) \propto m^H \Longleftrightarrow \ln \mathscr{R}_{Y}^{\mathrm{AL}}(m) \sim \gamma + H \cdot \ln m.
\end{equation}
for \eqref{RS-9}. 
It should be noted that the estimation for the Hurst exponent via revised statistic $\mathscr{R}_{Y_\tau}^{\mathrm{AL}}$ will be smaller than the true value when $H > 0.5$ if replacing the \eqref{RS-9} with \eqref{RS-11}.

\subsubsection{Algorithm for R/S Estimator}

The \Algr \ref{alg-RS}, which is designed according to \eqref{RS-1}, \eqref{RS-2}, \eqref{RS-3}, \eqref{RS-4}, \eqref{RS-5}, \eqref{RS-6}, \eqref{RS-7}, \eqref{RS-8}, \eqref{RS-9} and \eqref{RS-12},  provides a procedure for calculating the Hurst exponent of a time sequence with the help of \eqref{RS-12}. 

\begin{breakablealgorithm}
	\caption{Rescaled Range Analysis Estimator for Hurst exponent}\label{alg-RS}
	\begin{algorithmic}[1]
		\Require Time sequences data $ \vec{X} $, window size $ w $, indicator \cpvar{flag} for the optimization method in linear regression.
		\Ensure Hurst exponent of the sequence $ \vec{X} $.
		\Function{EstHurstRS}{$ \vec{X},w, \cpvar{flag} $}
		\State $ N \gets  \ProcName{GetLength}(\vec{X}) $;
		\State $ \scrd{N}{opt}\gets\ProcName{SearchOptSeqLen}(N,w) $;
		\State $ \vec{T}\gets \ProcName{GenSbpf}(\scrd{N}{opt}, w) $; 
		\State $ n\gets  \ProcName{GetLength}(\vec{T})$; 
		\State $ \vec{S}\gets\vec{0}\in\ES{R}{n}{1} $; // For the statistics
		\For{$ \cpvar{idx}\in\seq{1,2,\cdots,n} $}
		\State $ m \gets T_{\cpvar{idx}}$;
		\State $ k\gets\scrd{N}{opt}/{m} $;
		\State $ \vec{L}\gets\vec{0}\in\ES{R}{n}{1} $; // For the rescaled range
		\For{$ \tau\in\seq{1,2,\dots,k} $}
		\State $ E_\tau\gets\Ave{i}{1}{m}{X_{(\tau-1)m+i}} $; 
		\State $\vec{B}_\tau\gets \vec{0}\in \ES{R}{m}{1}$;
		\For{$j\in \seq{1, 2, \cdots, m}$}
		\State $ B^j_\tau =X_{(\tau-1)m+j}-E_\tau $;
		\EndFor
		\State $\vec{Y}_\tau\gets \vec{0}\in \ES{R}{m}{1}$;
		\For{$i\in \set{1, \cdots, m}$}
		\State $ Y^i_\tau\gets \OpCsum{j}{1}{i}{B^j_\tau}$; 
		\EndFor
		\State $ r_\tau(m)\gets\max\limits_{1\le i\le m}Y^i_\tau-\min\limits_{1\le i\le m}Y^i_\tau $;
		\State $ s_\tau(m)\gets \OpStd{i}{1}{m}{B^i_\tau} $;
		\State $ L_\tau \gets r_\tau(m)/s_\tau(m) $;
		\EndFor
		\State $ S_{\cpvar{idx}}  \gets \Ave{\tau}{1}{k}{L_\tau}$;// by \eqref{RS-5}
		\EndFor
		\State $ \seq{\mat{A},\vec{b}}\gets\ProcName{FormatPowLawData}(\vec{T},\vec{S},n) $;
		\State $\vec{p}\gets \ProcName{LinearRegrSolver}(\mat{A}, \vec{b}, n, \cpvar{flag})$;
		\State $\scrd{\beta}{RS} \gets p_2$ ; // $\vec{p} = \trsp{[\alpha,\beta]}$;
		\State $H \gets \scrd{\beta}{RS}$;
		\State\Return $H$;
		\EndFunction
	\end{algorithmic}
\end{breakablealgorithm}

\subsection{Triangles Total Areas (TTA) Method }

\subsubsection{Principle of TTA Estimator}

\begin{figure}[htbp]
	\centering
	\includegraphics[width=.5\textwidth]{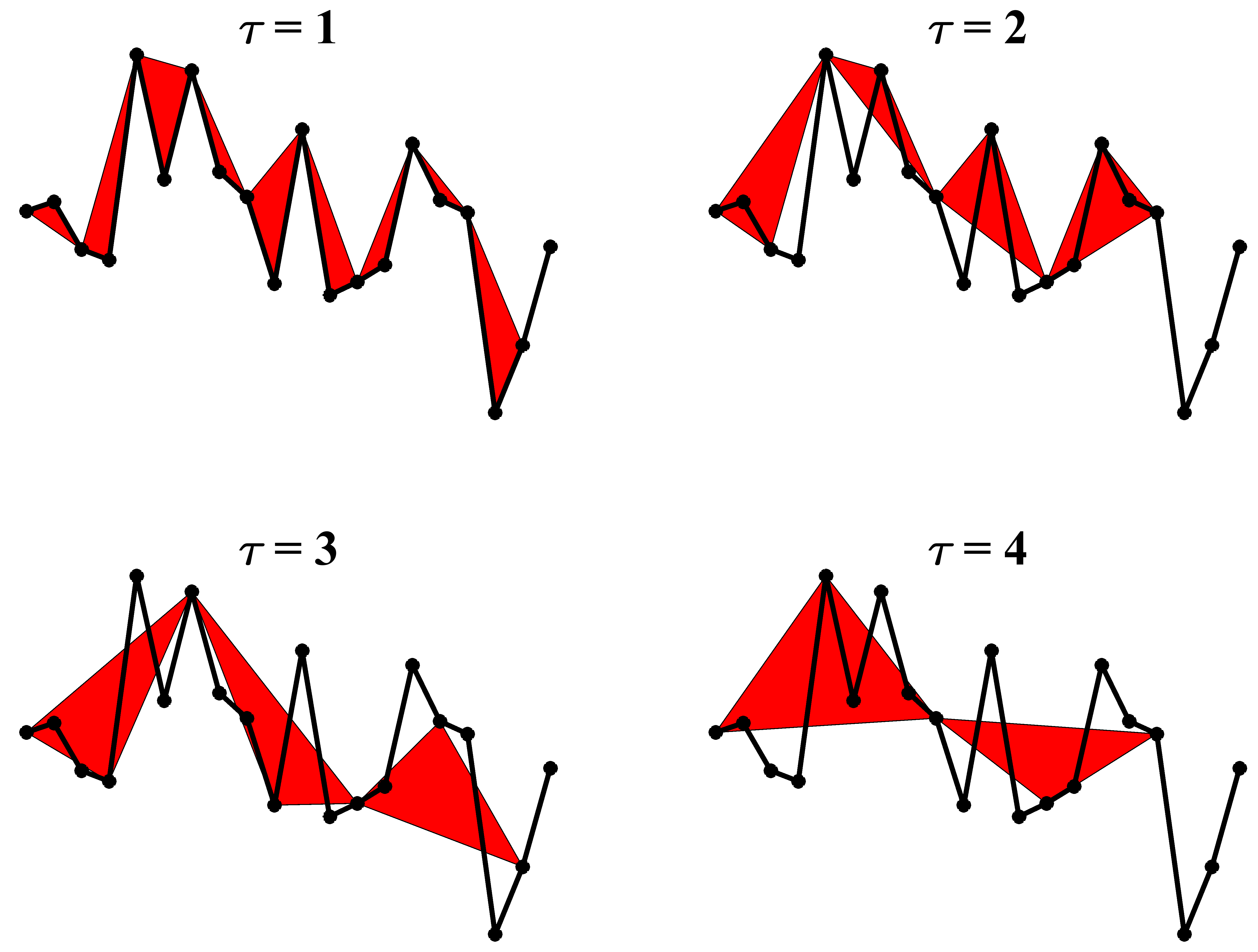}
	\caption{Construction of triangles with four different lags $ \tau=1,2,3,4 $.}\label{figure-TTA}
\end{figure}

For time sequence $ \set{X_t}_{t=1}^N $, we can derive the \textit{triangles total areas} (TTA) method with the cumulative sequence 
\begin{equation} \label{TTA-Yi}
Y_i = \OpCsum{t}{1}{i}{X_t - \mean{X}}, \quad 1\le i \le n.
\end{equation}
For the fixed time lag $\tau\in \mathbb{N}$ and $i$-th group of vertices 
$\set{P_i, Q_i, R_i}\subset \mathbb{R}^2$ such that
\begin{equation*}
\left\{
\begin{array}{ll}
P_i = (i, Y_i)  & \\
Q_i = (i+\tau,Y_{i+\tau}) & 1 \le i \le \mfloor{\frac{N-1}{2\tau}} \\
R_i = (i+2\tau,Y_{i+2\tau})
\end{array}
\right.
\end{equation*}
for the triangle $\Delta P_iQ_iR_i$, its  area can be calculated with the 3-order determinant, viz.
\begin{equation}\label{TTA-1}
\begin{aligned}
	A_i &=\dfrac{1}{2}\abs{\det{\left(\begin{matrix}
			i&i+\tau&i+2\tau\\
			Y_i&Y_{i+\tau}&Y_{i+2\tau}\\
			1&1&1
			\end{matrix}\right)}} \\
	&=\dfrac{\tau}{2}\abs{Y_{i+2\tau}-2Y_{i+\tau}+Y_{i}}, \quad 1 \le i \le \mfloor{\frac{N-1}{2\tau}}
\end{aligned}
\end{equation}
then the total area of the triangles is 
\begin{equation}\label{TTA-2}
	\scrd{A}{total}(\tau)=\dfrac{\tau}{2}\sum_{j=1}^{\mfloor{\frac{N-1}{2\tau}}}A_i.
\end{equation}
\Fig \ref{figure-TTA}  illustrates the relevant details of the construction for each triangle and the total area.
\\

Lotfalinezhad and Maleki \cite{Lotfalinezhad2020tta} showed that  
\begin{equation}\label{TTA-3}
\scrd{A}{total}(\tau)\propto \tau^H \Longleftrightarrow
\ln \scrd{A}{total}(\tau) \sim \scrd{\alpha}{TTA} + \scrd{\beta}{TTA}\cdot \ln \tau
\end{equation}
Consequently, we have
\begin{equation}\label{TTA-4}
\scrd{\hat{H}}{TTA} = \scrd{\hat{\beta}}{TTA}
\end{equation}
by linear regression. 
The \textit{triangles areas method} (TA), a modification of the TTA method,  proposed in \cite{Gomez2021theoretical}  has similar principle, and the modification is just to consider the distribution of the area of the triangles instead of the distribution of the sum of the total areas of the triangles involved.

\subsubsection{Algorithm for TTA Estimator}
With the help of \eqref{TTA-Yi}, \eqref{TTA-1}, \eqref{TTA-2}, \eqref{TTA-3} and \eqref{TTA-4}, we can desgin the estimation algorithm for the TTA method, please see \Algr alg-TTA.  Note that we set $ \scrd{\tau}{\max}=10 $ in \Algr \ref{alg-TTA} as we have done in \Algr \ref{alg-GHE}.

\begin{breakablealgorithm}
	\caption{Total Triangle Area Method for Estimating Hurst Exponent}\label{alg-TTA}
	\begin{algorithmic}[1]
		\Require Time series data $ \vec{X} $, indicator \cpvar{flag} for the optimization method in linear regression.
		\Ensure Hurst exponent of the sequence $ \vec{X} $.
		\Function{EstHurstTTA}{$ \vec{X}, \cpvar{flag}$}
		\State $ n\gets 10 $; // $ \scrd{\tau}{\max}=10 $
		\State $ N \gets  \ProcName{GetLength}(\vec{X}) $;
		\State $ \vec{T}\gets\seq{1,2,\cdots,n} $;
		\State $ \vec{S}\gets\vec{0}\in\ES{R}{n}{1} $; // For the statistics
		\State $ \mean{X}\gets\Ave{i}{1}{N}{X_i}$;
		\For{$i\in \seq{1,2,\cdots,N}$}
		\State $ Y_i\gets\OpCsum{j}{1}{i}{X_{j}-\mean{X}}$;
		\EndFor
		\For{$ \cpvar{idx}\in\seq{1,2,\cdots,n} $}
		\State $ \cpvar{sum}\gets 0 $;
		\For{$ i\in\seq{1,2,\cdots,\mfloor{\frac{N-1}{2\tau}}} $}
		\State $ j\gets2(i-1)\tau+1 $;// by \eqref{TTA-1}
		\State $ \cpvar{sum}\gets  \cpvar{sum}+\abs{Y_{j+2\tau}-2Y_{j+\tau}+Y_{j}} $; 
		\EndFor
		\State $ S_\cpvar{idx} \gets \cpvar{idx}\cdot \cpvar{sum}/2$;
		\EndFor
		\State $ \seq{\mat{A},\vec{b}}\gets\ProcName{FormatPowLawData}(\vec{T},\vec{S},n) $;
		\State $\vec{p}\gets \ProcName{LinearRegrSolver}(\mat{A}, \vec{b}, n, \cpvar{flag})$;
		\State $\scrd{\beta}{TTA} \gets p_2$ ; // $\vec{p} = \trsp{[\alpha,\beta]}$;
		\State $ H\gets \scrd{\beta}{TTA} $;// by \eqref{TTA-4}
		\State \Return $ H $;
		\EndFunction
	\end{algorithmic}
\end{breakablealgorithm}

\subsection{Periodogram Method (PM)}
\subsubsection{Principle of Periodogram Estimator}

Geweke and Porter-Hudak  \cite{Geweke1983estimation} proposed the \textit{periodogram method} (PM)  for estimating the Hurst exponent. The periodogram for a time sequence $ \set{X_t: 1\le t \le N}$ can be calculated  by
\begin{equation}\label{PM-1}
	I(k)=\dfrac{1}{N}\abs{\sum_{t=1}^{N}X_t \me^{-\frac{2\pi \mi}{N}  (k-1)(t-1)}}^2, \quad 1 \le k \le N
\end{equation}
where  $\mi = \sqrt{-1}$ and $ I(k)$ in \eqref{PM-1} is the squared absolute value of the DFT of the sequence $X_t$. Weron et al. showed that \cite{Weron2002estimating} 
\begin{equation}\label{PM-2}
I(k) \propto \left[4\textrm{sin}^2\left(\frac{k}{2N}\right)\right]^{\frac{1}{2}-H}, \quad 1\le k \le \mfloor{\frac{N}{2}}
\end{equation}
or equivalently
\begin{equation}\label{PM-3}
\ln I(k) \sim \scrd{\alpha}{PM} + \scrd{\beta}{PM}\cdot \ln \left[4\textrm{sin}^2\left(\frac{k}{2N}\right)\right], \quad 1\le k \le \mfloor{\frac{N}{2}}.
\end{equation}
In consequence, we have 
\begin{equation}\label{PM-4}
\scrd{\hat{H}}{PM} = \frac{1}{2} - \scrd{\hat{\beta}}{PM}
\end{equation}
by performing the linear regression for \eqref{PM-3}.

\subsubsection{Algorithm for Periodogram Estimator}

It should be noted that the spectrum-domain method relies more advanced mathematical concepts and tools. In the \Algr \ref{alg-PM} based on \eqref{PM-1}, \eqref{PM-2}, \eqref{PM-3} and \eqref{PM-4}, we can take the procedure \ProcName{FFT} to transform a time sequence into spectrum-domain. Since there are lots of toolboxes for  the \ProcName{FFT} in C/C++/MATLAB/Python/R, the details for the principle and algorithm implementation  are omitted here.

\begin{breakablealgorithm}
	\caption{Periodogram Estimator for Hurst exponent}\label{alg-PM}
	\begin{algorithmic}[1]
		\Require Time sequence $ \vec{X} $, the cut-off frequency $\scrd{f}{cutoff}$,  indicator \cpvar{flag} for the optimization method in linear regression.
		\Ensure Hurst exponent of the sequence $ \vec{X} $.
		\Function{EstHurstPeriodDiagram}{$ \vec{X}, \scrd{f}{cutoff},\cpvar{flag} $}
		\State $ N \gets  \ProcName{GetLength}(\vec{X}) $;% // Length of original sequence
		\State $ \vec{Y} \gets $\ProcName{FFT}$ (\vec{X}) $; 
		%\State $ \vec{Y} \gets \vec{Y}(2:N)$; // Remove first element
		\State $ \vec{T}\gets \emptyset$; // $\vec{T}$ is used for the $4\sin^2(k/(2N))$, according to \eqref{PM-2}
		\State $\vec{S} \gets \emptyset$; // $\vec{S}$ is used for the periodogram $I(k)$, according to \eqref{PM-1}
		\For{$ k \in \seq{2,\cdots,\mfloor{\frac{N}{2}}}$} \quad// Attention, please!
		\State $ f\gets k/N $; \quad // Calculate frequencies
		\If{$ f \le \scrd{f}{cutoff}$}
		\State $\vec{T} \gets \vec{T} \cup \set{4\sin^2(f/2)} $; 
		\State $ \vec{S} \gets \vec{S} \cup 
		\set{\abs{Y_k}^2/N}$;
		\EndIf
		\EndFor
		\State $n\gets \ProcName{GetLength}(\vec{T})$;
		\State $ \seq{\mat{A},\vec{b}}\gets\ProcName{FormatPowLawData}(\vec{T},\vec{S},n) $;
		\State $\vec{p}\gets \ProcName{LinearRegrSolver}(\mat{A}, \vec{b}, n, \cpvar{flag})$;
		\State $\scrd{\beta}{PD} \gets p_2$; \quad // $\vec{p} = \trsp{[\alpha,\beta]}$;
		\State $ H \gets 0.5-\scrd{\beta}{PD} $;// by \eqref{PM-4}
		\State\Return $ H $;
		\EndFunction
	\end{algorithmic}
\end{breakablealgorithm}

\subsection{Discrete Wavelet Transform (DWT) Method}

It is also feasible to estimate the Hurst exponent with discrete wavelet transform.  The time sequence $ \vec{X} = \set{X_t}^N_{t=1}$ can be transformed into the spectrum-domain by
\begin{equation*}
W_{\vec{X}}(a,b) = \mathrm{DWT}^a_b(\vec{X}, \psi)
\end{equation*}
where $ a $ is the scale parameter, $b$ is the location parameter, $ \psi $ is the wavelet function and $\mathrm{DWT}^a_b$ is the DWT for the details of the time sequence according to \eqref{eq-DWT}. For the given scale $ a $, we can find a representation of the wavelet ``energy'' or amplitude and study its scaling for exploring the power law of interest when estimating the Hurst exponent\cite{Abry1998wavelet}. 

\subsubsection{Average Wavelet Coefficient Method  (AWC)}

The \textit{average wavelet coefficient} (AWC) method is based on the self-affine correlations of the DWT of a time sequence, which can be used for estimating the Hurst exponent. Simonsen \cite{Simonsen1998determination} showed that 
\begin{equation}\label{AWC-1}
	\scru{W}{awc}_{\vec{X}}(a)=\frac{1}{\card{I_a}}\sum_{b\in I_a} \abs{W(a,b)}\sim a^{H-0.5}
\end{equation}
where $ I_a $ is defined by \eqref{eq-Ib}. Taking the logarithms of both sides in \eqref{AWC-1}, we immediately have
\begin{equation}\label{AWC-2}
\ln \scru{W}{awc}_{\vec{X}}(a) \sim \scrd{\alpha}{AWC} + \scrd{\beta}{AWC} \cdot \ln a. 
\end{equation}
Consequently, we can obtain
\begin{equation}\label{AWC-3}
 \scrd{\hat{H}}{AWC} =  \scrd{\hat{\beta}}{AWC} + \frac{1}{2}
\end{equation}
by \eqref{AWC-2} with the help of linear regression. 

\subsubsection{Variance Versus Level (VVL) Method }

Similar to the AWC method, we can construct the \textit{variance versus level} VVL spectrum over all of the location parameters $ b\in I_a$ for the given scale $a$. Let 
\begin{equation}\label{VVL-1}
\scru{W}{vvl}_{\vec{X}}(a) = \frac{1}{\card{I_a}-1}\sum_{b\in I_a} 
\left[|W(a,b)| - \scru{W}{awc}_{\vec{X}}(a) \right]^2
\end{equation}
be the variance of $|W(a,b)|$ respect to the discrete location variable $b$.
Flandrin \cite{Flandrin1992wavelet} discovered the asymptotic formula  
\begin{equation}\label{VVL-2}
\scru{W}{vvl}_{\vec{X}}(a)\sim a^{2H-1}
\end{equation}
for \eqref{VVL-1}. Equivalently, we have
\begin{equation}\label{VVL-3}
\ln \scru{W}{vvl}_{\vec{X}}(a) \sim \scrd{\alpha}{VVL} + \scrd{\beta}{VVL} \cdot \ln a
\end{equation}
by taking the logarithm on the both sides of \eqref{VVL-2}. In consequence, the new estimator of Hurst exponent can be written by
 \begin{equation}\label{VVL-4}
\scrd{\hat{H}}{VVL} = (1+ \scrd{\hat{\beta}}{VVL})/2.
\end{equation}

\subsubsection{Algorithm for DWT Estimator}

It is easy to find there is a unified formula for the AWC and VVL methods. Combining \eqref{AWC-3} and \eqref{VVL-4}, we have
\begin{equation}\label{DWT-H}
\begin{aligned}
\scrd{\hat{H}}{DWT} 
&= \frac{1}{2} + \frac{\scrd{\hat{\beta}}{DWT}}{r}  \\
&=\left\{
\begin{array}{ll}
0.5 + \scrd{\hat{\beta}}{AWC}  & r = 1\\
0.5 + 0.5\cdot \scrd{\hat{\beta}}{VVL}, & r= 2
\end{array}
\right.
\end{aligned}
\end{equation}
Thus it is convenient for us to design a unified interface for estimating the Hurst exponent with the AWC method and the VVL method. 

The \Algr \ref{alg-DWT} is designed according to \eqref{VVL-1}, \eqref{VVL-2}, \eqref{VVL-3},\eqref{VVL-4} and \eqref{DWT-H} with the help of DWT. Currently, the DWT algorithm is a built-in function in MATLAB, Python and R. We remark that here we take a unified interface \ProcName{Wavedec} to achieve the multilevel DWT for 1-dim time sequences.

\begin{breakablealgorithm}
	\caption{Discrete Wavelet Transform Estimator for Hurst exponent}\label{alg-DWT}
	\begin{algorithmic}[1]
		\Require Time sequences data $ \vec{X} $, integer $ r\in\seq{1,2} $ for the AWC/VVL method, indicator \cpvar{flag} for the optimization method in linear regression.
		\Ensure Hurst exponent of the sequence $ \vec{X} $.
		\Function{EstHurstDWT}{$ \vec{X}, r, \cpvar{flag} $}
		\State $ N \gets  \ProcName{GetLength}(\vec{X}) $;% // Length of original sequence
		\State $ n\gets\mfloor{  \log_2 (N)} $; // Calculate appropriate decomposition level
		\If{$ r=1 $}
			\State $ \vec{W}\gets\ProcName{\textrm{Wavedec}}(\vec{X},\lstinline|"db24"|,n) $; // 24-th order Daubechies DWT
		\Else
			\State $ \vec{W}\gets\ProcName{\textrm{Wavedec}}(\vec{X},\lstinline|"haar"|,n) $; // Haar DWT
		\EndIf
		\State $ \vec{T}\gets\vec{0}\in\ES{R}{n}{1} $; // For the scale
		\State $ \vec{S}\gets\vec{0}\in\ES{R}{n}{1} $; // For the AWC-spectrum
		\For{$ \cpvar{idx}\in\seq{1,2,\dots,n} $}
		\State $ T_\cpvar{idx}\gets{2^\cpvar{idx}} $; // Add scale coefficient
		\State $ \scrd{\vec{L}}{pos}\gets \abs{W_\cpvar{idx}} $; // All location parameters corresponding to scale $ 2^\cpvar{idx} $
		\If{$ r=1 $}
			\State $ S_\cpvar{idx} \gets \ProcName{Mean}(\scrd{\vec{L}}{pos})$;// AWC-spectrum, by \eqref{AWC-1}
		\Else
			\State $ S_{\cpvar{idx}}  \gets \Var(\scrd{\vec{L}}{pos})$;// VVL-spectrum, by \eqref{VVL-1}
		\EndIf
		\EndFor
		\State $ \seq{\mat{A},\vec{b}}\gets\ProcName{FormatPowLawData}(\vec{T},\vec{S},n) $;
		\State $\vec{p}\gets \ProcName{LinearRegrSolver}(\mat{A}, \vec{b}, n, \cpvar{flag})$;
		\State $\scrd{\beta}{DWT} \gets p_2$ ; // $\vec{p} = \trsp{[\alpha,\beta]}$;
		\State $H\gets \scrd{\beta}{DWT}/r+0.5 $;// by \eqref{DWT-H}
		\State\Return $ H $;
		\EndFunction
	\end{algorithmic}
\end{breakablealgorithm}

\subsection{Local Whittle (LW) Method }

\subsubsection{Principle of LW Estimator}

As the similar process stated in the PM method, for the vector 
$\vec{\lambda} = \trsp{\left[\lambda_1, \cdots, \lambda_n\right]}$ 
such that $\lambda_j = \frac{2\pi j}{N}$ for $j\in \set{1, 2, \cdots, 
 n = \mfloor{\frac{N}{2}}}$, we set  
\begin{equation}\label{Whittle-4}
\begin{aligned}
\vec{I}(\vec{\lambda})
= \trsp{[I_1, \cdots, I_n]}
=\trsp{\left[\abs{\omega(\lambda_1)}^2, \cdots, \abs{\omega(\lambda_n)}^2\right]} 
\end{aligned}
\end{equation}
where \begin{equation}\label{Whittle-3}
\omega(\lambda_j) =\frac{1}{N} \sum_{t=1}^{N}X_te^{\mi (t-1)\lambda_j}
\end{equation}
is the DFT of the TS $X_t=\set{X_1, \cdots, X_N}$. Kunsch et al \cite{Kunsch2020statistical} showed that the Hurst exponent can be estimated by solving the following optimization problem
\begin{equation}\label{Whittle-5}
\scrd{\hat{H}}{LW} = \arg \min\limits_{H\in(0,1)}  \psi(H)
\end{equation}
with the objective function 
\begin{equation}\label{Whittle-6}
\psi(H) = \ln\left[ \dfrac{1}{n}\sum_{j=1}^{n}\lambda^{2H-1}_jI_j\right] - \dfrac{2H-1}{n}\sum_{j=1}^{n}\ln \lambda_j
\end{equation}
where the $j$-th component $I_j$ is defined by \eqref{Whittle-4}.
For more details, please see Robinson's work \cite{Robinson1995LocalWhittle}.

\subsubsection{Algorithm for LW Estimator}

The procedure \ProcName{EstHurstLW} listed in \Algr \ref{alg-LocalWhittle} is designed for estimating the Hurst exponent with the LW method specified by \eqref{Whittle-4},
 \eqref{Whittle-3}, \eqref{Whittle-5} and \eqref{Whittle-6}. We remarked that there are two procedures that are involved in \Algr \ref{alg-LocalWhittle}:
\begin{itemize}
\item the procedure \ProcName{ObjFunLW} is used to compute the value of $\psi(H)$ determined by \eqref{Whittle-6}, and 
\item the procedure \ProcName{LocMinSolver} to find the minimum of $\psi(H)$, please see the \Algr \ref{alg-localmin} in the Appendix \ref{sec-appendix} for more details. 
\end{itemize}

\begin{breakablealgorithm}
	\caption{Estimating the Hurst exponent with the Local Whittle method}
	\label{alg-LocalWhittle}
	\begin{algorithmic}[1]
		\Require Time sequence $ \vec{X} $.
		\Ensure Hurst exponent of the sequence $ \vec{X} $.
		\Function{EstHurstLW}{$ \vec{X} $}
		\State $ N \gets  \ProcName{GetLength}(\vec{X}) $; // Length of the original sequence
		\State $ \vec{Y} \gets $\ProcName{FFT}$ (\vec{X}) $; 
		\State $ n\gets\mfloor{N/2} $;
		\State $ \vec{T}\gets\vec{0}\in\ES{R}{n}{1} $; // for the frequencies
		\State $ \vec{S}\gets\vec{0}\in\ES{R}{n}{1} $; // for the periodogram
		\For{$ \cpvar{idx} \in \seq{1,\cdots,n} $} \quad // Attention, please!
		\State $ T_\cpvar{idx}\gets{\cpvar{idx}/N} $; // Calculate frequencies
		\State $ S_\cpvar{idx} \gets {\abs{Y_{\cpvar{idx}+1}}^2} $; // Periodogram, see \eqref{Whittle-4}
		\EndFor
		\State $ H \gets\ProcName{LocMinSolver}(\ProcName{ObjFunLW},0.001, 0.999, 10^{-8}, \vec{T}, \vec{S})$; \ // by \eqref{Whittle-5}
		\State\Return $ H $;
		\EndFunction
	\end{algorithmic}
\end{breakablealgorithm}

\begin{breakablealgorithm}
	\caption{Computing the objective function $\psi(H)$ in Local Whittle method}\label{alg-LocalWhittle-target}
	\begin{algorithmic}[1]
		\Require Variable  $ H \in (0,1) $, frequency vector $\vec{T}$ and periodogram vector $\vec{S}$.
		\Ensure $\psi(H)$.
		\Function{ObjFunLW}{$x,\seq{\vec{T}, \vec{S}} $}
		\State $ n\gets \ProcName{GetLength}(\vec{T}) $; \quad // $n = \mfloor{N/2}$ 
		\State $ y\gets\ln\left(\dfrac{1}{n}\sum\limits_{i=1}^{n}T_i^{2H-1}S_i\right)-\dfrac{2H-1}{n}\sum\limits_{i=1}^{n}\ln T_i $;
		\State\Return $ y $; \quad // by \eqref{Whittle-6}
		\EndFunction
	\end{algorithmic}
\end{breakablealgorithm}

\subsection{Least Squares via Standard Deviation (LSSD)}

\subsubsection{Principle of LSSD Estimator}

The Hurst exponent can also be estimated with the \textit{least squares via standard deviation} (LSSD). The steps are summarized as follows:
\begin{itemize}
\item[0)] Pre-conditioning: Dividing the sequence $\set{X_t}_{t=1}^N$ into $k=\mfloor{N/m}$ subsequences with the same size $m$ according to \eqref{eq-seq-partition} 
where $ m\in \set{1, 2,\cdots,\scrd{m}{max}}$ 
such that $\scrd{m}{max}\ge\mfloor{\frac{N}{10}}$ 
and each $m$ corresponds to a partition scheme.

\item[i)] Calculating the cumulative sum of the $i$-th subsequence $X_{(i)} = \set{X_{(i-1)m+j}: 1\le j \le m}$, viz.
\begin{equation} \label{LSSD-1}
Z^m_i = \OpCsum{j}{1}{m}{X_{(i-1)m+j}}, \quad 1\le i\le \mfloor{\frac{N}{m}}
\end{equation}
\item[ii)] Creating the standard deviation sequences $\set{s_m}_{m=1}^{\scrd{m}{max}}$ by
\begin{equation} \label{LSSD-2}
s_m = \OpStd{i}{1}{m}{Z^m_i}, \quad 1\le m \le \scrd{m}{max}
\end{equation}
Suppose
\begin{equation*}
\Expt{\mean{s_m}} = \Expt{\Ave{m}{1}{\scrd{m}{max}}{s_m}} = \sigma,
\end{equation*}
then  the self-similarity property of the sequence implies that \cite{Koutsoyiannis2003climate}
 \begin{equation*}
 \Expt{s_m} \approx \sigma \cdot \scrd{c}{LSSD}(m,H) \cdot m^H
 \end{equation*}
 where $H$ is the Hurst exponent and
\begin{equation}\label{LSSD-3}
	\scrd{c}{LSSD}(m,H)=\sqrt{\dfrac{N/m-(N/m)^{2H-1}}{
	 N/m - 1/2}}.
\end{equation}
\item[iii)] Constructing the optimization problem. Koutsoyiannis et al. 
 \cite{Koutsoyiannis2003climate-supple} introduced the following function for fitting error 
\begin{equation}\label{LSSD-4}
\begin{aligned}
	&\scrd{\mathcal{E}}{LSSD}^2(\sigma,H)\\
	&= \sum^{\scrd{m}{max}}_{m=1} \frac{\left[\ln  \dfrac{\Expt{s_m}}{s_m}\right]^2}{m^p}+ \frac{H^{q+1}}{q+1} \\
	&=\sum_{m=1}^{\scrd{m}{max}}\dfrac{\left[\ln\sigma+H\cdot\ln m+\ln \scrd{c}{LSSD}(m,H)-\ln s_m\right]^2}{m^p} \\
	&\quad +\dfrac{H^{q+1}}{q+1}
\end{aligned}
\end{equation}
where $p\in \set{0, 1, 2, \cdots}$ is a weight parameter and $ H^{q+1}/(q+1) $ is  a penalty factor with default value $ q=50$. 
\end{itemize}
Let
\begin{equation}
\left\{
\begin{aligned}
	a_{11}  &=\sum_{m=1}^{\scrd{m}{max}}\dfrac{1}{m^p} \\
	a_{12}  &=\sum_{m=1}^{\scrd{m}{max}}\dfrac{\ln m}{m^p}\\ 
	d_m(H) &=\ln m+\dfrac{\ln(N/m)}{1-(N/m)^{2-2H}}
\end{aligned}
\right.
\end{equation}
and 
\begin{equation}
\left\{
\begin{aligned}
	a_{21}(H)  &=\sum_{m=1}^{\scrd{m}{max}}\dfrac{d_m(H)}{m^p},\\
	a_{22}(H)  &=\sum_{m=1}^{\scrd{m}{max}}\dfrac{d_m(H)\ln m}{m^p}, \\
	b_1(H) &=\sum_{m=1}^{\scrd{m}{max}}\dfrac{\left[\ln s_m - \ln \scrd{c}{LSSD}(m,H)\right]}{m^p}, \\
	b_2(H) &=\sum_{m=1}^{\scrd{m}{max}}\dfrac{d_m(H) \left[\ln s_m - \ln \scrd{c}{LSSD}(m,H)\right]}{m^p}.	
\end{aligned}
\right.
\end{equation}
With the help of the least squares approach, we can solve \eqref{LSSD-4} to obtain a fixed-point equation for the Hurst exponent, which can be written by 
\begin{equation}\label{LSSD-5}
H=\scrd{\Phi}{LSSD}(H)
\end{equation}
in which 
\begin{equation}\label{LSSD-6}
	\scrd{\Phi}{LSSD}(H) = \dfrac{a_{11}[b_2(H)-H^q]-a_{21}(H)b_1(H)}{a_{11}a_{22}(H)-a_{21}(H)a_{12}}.
\end{equation}
Obviously, we can use the Newton's iterative method 
or direct iterative method to solve the fixed-point in order to estimate the exponent. Koutsoyiannis et al. \cite{Koutsoyiannis2003climate-supple} pointed out that there is a unique fixed-point for the equation \eqref{LSSD-5}, which can be solved with \Algr \ref{alg-fixed-point}. 
For more details of fixed-point algorithm, please see Zhang et al. \cite{ZhangHY2024Kuiper} or the toolbox of MATLAB, Python, and so on. 

\subsubsection{Algorithm for LSSD Estimator}

The procedure \ProcName{EstHurstLSSD}  listed in \Algr \ref{alg-LSSD} 
is designed to estimate the Hurst exponent with the LSSD method based on \eqref{LSSD-1}, \eqref{LSSD-2}, \eqref{LSSD-3}, \eqref{LSSD-4}, \eqref{LSSD-5} and \eqref{LSSD-6}. 
Note that the procedure \ProcName{CtmLSSD} is used to compute the contractive mapping  $\scrd{\Phi}{LSSD}(H)$ and the procedure \ProcName{FixedPointSolver} listed in \Algr \ref{alg-fixed-point}  provides a general interface for solving the  fixed-point of some nonlinear equation.

\begin{breakablealgorithm}
	\caption{LSSD Estimator}\label{alg-LSSD}
	\begin{algorithmic}[1]
		\Require Time sequence $ \vec{X} $, weight $ p $, penalty parameter $ q $, precision $ \epsilon $ with default value $\epsilon = 10^{-4}$.
		\Ensure Hurst exponent of the sequence $ \vec{X} $.
		\Function{EstHurstLSSD}{$ \vec{X},p,q,\epsilon $}
		\State $ N \gets  \ProcName{GetLength}(\vec{X}) $;% // Length of original sequence
		\State $ \scrd{m}{max}\gets\mfloor{N/10} $;
		\State $ \vec{T}\gets\seq{1,2,\cdots,\scrd{m}{max}} $;
		\State $ \vec{S}\gets\vec{0}\in\ES{R}{\scrd{m}{max}}{1} $; // For the standard deviation
		\For{$ \cpvar{idx}\in\seq{1,2,\cdots,\scrd{m}{max}} $}
		\State $ m\gets\cpvar{idx} $; 
		\State $ k\gets\mfloor{N/m} $; 
		\State $\vec{Z}\gets \vec{0}\in \ES{R}{k}{1}$; 
		\For{$i\in\seq{1, 2, \cdots, k}$}
		\State $ Z_i\gets\sum\limits_{j=1}^{m}X_{(i-1)m+j}$; // compute $Z^m_i$ by \eqref{LSSD-1}
		\EndFor
		\State $ S_\cpvar{idx} \gets \OpStd{i}{1}{k}{Z_i}$;
		\EndFor
		\State $ H \gets\ProcName{FixedPointSolver}(\ProcName{CtmLSSD},0.5,\epsilon, N, p, q, \vec{T}, \vec{S}) $; // by \eqref{LSSD-5}
		\State\Return $ H $;
		\EndFunction
	\end{algorithmic}
\end{breakablealgorithm}
Please note that the Line 11 in \Algr \ref{alg-LSSD} is based on \eqref{LSSD-1}. 
The procedures \ProcName{FunCmLSSD} listed in \Algr \ref{alg-cmH} and \ProcName{FunDmH} listed in \Algr \ref{alg-dmH} are used to compute the $\scrd{c}{LSSD}(m,H)$ and $d_m(H)$ respectively. Furthermore, the procedure \ProcName{CtmLSSD}
for computing the $\scrd{\Phi}{LSSD}(H)$ listed in \Algr \ref{alg-ctm-LSSD} is an argument of high order procedure \ProcName{FixedPointSolver}.

\begin{breakablealgorithm}
\caption{Computing the $\scrd{c}{LSSD}(m,H)$} \label{alg-cmH}
\begin{algorithmic}[1]
\Require Positive integer $m\in \set{1, 2, \cdots, \scrd{m}{max}}$, parameter $H\in (0,1)$, Positive integer $N$
\Ensure The value of $\scrd{c}{LSSD}(m,H)$ 
\Function{FunCmLSSD}{$m, N, H$}
\State $u \gets N/m$;
\State $c \gets \sqrt{(u-u^{2H-1})/(u-0.5)}$; // by \eqref{LSSD-3}
\State \Return $c$; 
\EndFunction
\end{algorithmic}
\end{breakablealgorithm}

\begin{breakablealgorithm}
\caption{Computing the $d_m(H)$} \label{alg-dmH}
\begin{algorithmic}[1]
\Require Positive integer $m\in \set{1, 2, \cdots, \scrd{m}{max}}$, parameter $H\in (0,1)$, Positive integer $N$
\Ensure The value of $d_m(H)$ 
\Function{FunDmH}{$m, N, H$}
\State $u \gets N/m$;
\State $d \gets \ln m + \ln u/(1-u^{2-2H})$;
\State \Return $d$;
\EndFunction
\end{algorithmic}
\end{breakablealgorithm}

\begin{breakablealgorithm}
	\caption{Contractive Mapping for the  LSSD method}\label{alg-ctm-LSSD}
	\begin{algorithmic}[1]
		\Require Parameter $ H\in (0,1) $, length $ N $, weight $ p $, penalty parameter $ q $, scale  vector $ \vec{T} = \trsp{[1, 2, \cdots, \scrd{m}{max}]}$, standard deviation  vector $\vec{S} = \trsp{[s_1, s_2, \cdots, s_{\scrd{m}{max}}]}$.
		\Ensure The value of $\scrd{\Phi}{LSSD}(H)$.
		\Function{CtmLSSD}{$H,\seq{N, p, q, \vec{T}, \vec{S}} $}
		\State $ \scrd{m}{max}\gets\ProcName{GetLength}(\vec{T})$;
		\State $a_{11}\gets 0, a_{12}\gets 0$;
		\State $a_{21}\gets 0, a_{22}\gets 0$;
		\State $b_1\gets 0, b_2\gets 0$;		
		\For{$\cpvar{idx} \in\seq{1, 2, \cdots, \scrd{m}{max}}$}
		\State $m\gets T_{\cpvar{idx}}$;
		\State $s_m \gets S_{\cpvar{idx}}$;
		\State $c_m \gets \ProcName{FunCmLSSD}(m, N, H)$;
		\State $d_m \gets \ProcName{FunDmH}(m, N, H)$;
		\State $u\gets m^p$;
		\State $a_{11} \gets a_{11} + 1.0/u$; // by \eqref{LSSD-6}
		\State $a_{12} \gets a_{12} + \ln m /u$;
		\State $a_{21} \gets a_{21} + d_m/u$;
		\State $a_{22} \gets a_{22} + d_m\cdot \ln m / u$;
		\State $b_1 \gets b_1 + (\ln s_m - \ln c_m)/u$;
		\State $b_2 \gets b_2 + d_m\cdot(\ln s_m - \ln c_m)/u$;
		\EndFor
		\State $ g\gets\dfrac{a_{11}\cdot (b_2-H^q)-a_{21}\cdot b_1}{a_{11}\cdot a_{22}-a_{21}\cdot a_{12}} $; // by \eqref{LSSD-6}
		\State\Return $ g $;
		\EndFunction
	\end{algorithmic}
\end{breakablealgorithm}

\subsection{Least Squares via Variance (LSV)}

\subsubsection{Principle of LSV Estimator}

Similar to the LSSD-method, we can construct the variance sequences according to \eqref{LSSD-1}:
\begin{equation*}\label{LSV-1}
s^2_m = \left(\OpStd{i}{1}{m}{Z^m_i}\right)^2, \quad 1\le m \le \scrd{m}{max}
\end{equation*}
The self-similarity property implies that
\begin{equation*}
\Expt{s^2_m}=\scrd{c}{LSV}(m,H)\cdot m^{2H}\cdot \sigma^2
\end{equation*}
where
\begin{equation}\label{LSV-2}
	\scrd{c}{LSV}(m,H) 
= \dfrac{N/m-(N/m)^{2H-1}}{N/m-1} \\
\end{equation}
Tyralis et al. \cite{Tyralis2011simultaneous} introduced the  fitting error function
\begin{equation}\label{LSV-3}
	\scrd{\mathcal{E}}{LSV}^2(\sigma,H) 
	=\sum_{k=1}^{\scrd{k}{max}}\dfrac{\left[\scrd{c}{LSV}(m,H)k^{2H}\sigma^2-s^2_m\right]^2}{m^p}+\dfrac{H^{q+1}}{q+1}
\end{equation}
where $p\in \set{0,1, 2, \cdots}$ is a weight parameter  and  $ H^{q+1}/(q+1) $ is a penalty factor with default value $q=50$. 
By solving \eqref{LSV-3} with the least squares approach, Tyralis et al. \cite{Tyralis2011simultaneous} showed that the Hurst exponent can be estimated by solving the following optimization problem   
\begin{equation}\label{LSV-4}
\scrd{\hat{H}}{LSV} = \arg \min_{H\in (0,1)} \scrd{\Phi}{LSV}(H) 
\end{equation}
where
\begin{equation}\label{LSV-5}
\scrd{\Phi}{LSV}(H) =\sum_{m=1}^{\scrd{m}{max}}\dfrac{s^4_m}{m^p}-\dfrac{a^2_{12}(H)}{a_{11}(H)}+\dfrac{H^{q+1}}{q+1}
\end{equation}
in which
\begin{equation}\label{LSV-6}
\left\{\begin{aligned}
a_{11}(H)&=\sum_{m=1}^{\scrd{m}{max}}\dfrac{[\scrd{c}{LSV}(m,H)]^2 \cdot m^{4H}}{m^p}\\ 
a_{12}(H)&=\sum_{m=1}^{\scrd{m}{max}}\dfrac{\scrd{c}{LSV}(m,H)\cdot m^{2H}\cdot s^2_m}{m^p}
\end{aligned}\right.
\end{equation}

\subsubsection{Algorithm for LSV Estimator}

The procedure \ProcName{EstHurstLSV} listed in \Algr \ref{alg-LSV} is designed for estimating the Hurst exponent with the LSV method. 
For the high order procedure \ProcName{LocMinSolver} involved in \ProcName{EstHurstLSV}, please see \Algr \ref{alg-localmin}. Note that
the procedure \ProcName{ObjFunLSV}, the first argument of \ProcName{LocMinSolver}, is given in \Algr \ref{alg-obj-fun-LSV}.
\begin{breakablealgorithm}
	\caption{Estimating the Hurst exponent with the LSV method}
	\label{alg-LSV}
	\begin{algorithmic}[1]
		\Require Time sequence $ \vec{X} $, weight $ p\in\set{0 ,1, 2, \cdots}$, penalty parameter $ q\in \mathbb{N}$ with default value $q=50$, precision $ \epsilon $ with default value $\epsilon = 10^{-4}$.
		\Ensure Hurst exponent of the sequence $ \vec{X} $.
		\Function{EstHurstLSV}{$ \vec{X},p,q,\epsilon $}
		\State $ N \gets  \ProcName{GetLength}(\vec{X}) $;% // Length of original sequence
		\State $ \scrd{m}{max}\gets\mfloor{N/10} $;
		\State $ \vec{T}\gets\seq{1,2,\cdots,\scrd{m}{max}} $;
		\State $ \vec{S}\gets\vec{0}\in\ES{R}{\scrd{m}{max}}{1} $; // For the standard deviation
		\For{$ \cpvar{idx}\in\seq{1,2,\cdots, \scrd{m}{max}} $}
		\State $ m\gets\cpvar{idx} $; // size of the subsequences
		\State $ k\gets\mfloor{N/m} $; // number of subsequences
		\State $\vec{Z}\gets \vec{0}\in \ES{R}{k}{1}$;
		\For{$i\in \seq{1, 2, \cdots, k}$}
		\State $ Z_i\gets \sum\limits_{j=1}^{m}X_{(i-1)m+j} $;
		\EndFor
		\State $ S_\cpvar{idx}\gets\OpStd{i}{1}{m}{Z_i}$;
		\EndFor
		\State $ H \gets\ProcName{LocMinSolver}(\ProcName{ObjFunLSV},\ [0.001, 0.999],\epsilon, N, p, q, \vec{T}, \vec{S}) $; // by \eqref{LSV-4}
		\State\Return $ H $;
		\EndFunction
	\end{algorithmic}
\end{breakablealgorithm}

The procedures \ProcName{FunCmLSV} listed in \Algr \ref{alg-cm-LSV} is used to compute the $\scrd{c}{LSV}(m,H)$ according to \eqref{LSV-2}. Note that the \eqref{LSV-4} are used to compute the fixed-point for the Hurst exponent.

\begin{breakablealgorithm}
\caption{Computing the $\scrd{c}{LSV}(m,H)$} \label{alg-cm-LSV}
\begin{algorithmic}[1]
\Require Positive integer $m\in \set{1, 2, \cdots, \scrd{m}{max}}$, parameter $H\in (0,1)$, Positive integer $N$
\Ensure The value of $\scrd{c}{LSV}(m,H)$ 
\Function{FunCmLSV}{$m, N, H$}
\State $u \gets N/m$;
\State $c \gets (u-u^{2H-1})/(u-1)$; // by \eqref{LSV-2}
\State \Return $c$; 
\EndFunction
\end{algorithmic}
\end{breakablealgorithm}

The procedure \ProcName{ObjFunLSV} is used to compute the objective function $\scrd{\Phi}{LSV}(H)$ in \eqref{LSV-5} and \eqref{LSV-6} for estimating the Hurst exponent.
\begin{breakablealgorithm}
	\caption{Objective function $\scrd{\Phi}{LSV}(H)$ for the LSV method}\label{alg-obj-fun-LSV}
	\begin{algorithmic}[1]
		\Require Hurst exponent $H$, length $ N $, weight $ p $, penalty parameter $ q $, scale vector $\vec{T}=\trsp{[1, 2, \cdots, \scrd{m}{max}]}$ and standard deviation vector $\vec{S}=\trsp{[s_1, s_2, \cdots, \scrd{s}{max}]}$.
		\Ensure The value of $\scrd{\Phi}{LSV}(H)$ for the LSV method
		\Function{ObjFunLSV}{$H, N, p, q, \vec{T}, \vec{S} $}
		\State $ \scrd{m}{max}\gets\ProcName{GetLength}(\vec{T}) $;
		\State $b_2\gets\dfrac{H^{q+1}}{q+1}$;
		\State $a_{11}\gets 0, a_{12}\gets 0, b_1\gets 0$;
		\For{$\cpvar{idx}\in \set{1, 2, \cdots, \scrd{m}{max}}$}
		\State $m\gets T_{\cpvar{idx}}$;
		\State $s_m \gets \scrd{S}{idx}$;
		\State $c_m \gets \ProcName{FunCmLSV}(m, N, H)$;
		\State $u \gets m^p$;
		\State $b_1 \gets b_1 + s_m^4/u$;// by \eqref{LSV-6}
		\State $a_{11} \gets a_{11} + c_m^2\cdot m^{4H}/u$;
		\State $a_{12} \gets a_{12} + c_m \cdot m^{2H} \cdot s_m^2/u$;
		\EndFor
		\State $ g\gets b_1-\dfrac{a_{12}\cdot a_{12}}{a_{11}} + b_2 $;// by \eqref{LSV-5}
		\State \Return $ g $;
		\EndFunction
	\end{algorithmic}
\end{breakablealgorithm}

\section{Performance Assessment for the Typical Algorithms}
\label{sec-V-and-V}

\subsection{Experimental Setup}
Our experimental setup has the following configuration: Operating system
--- Ubuntu 22.04.3 LTS GNU/Linux(64-bit); Memory --- 32GB RAM; Processor --- 12th Gen Intel$ ^\circledR $ Cor$^\text{\scriptsize TM} $ i7-12700KF $ \times $ 20; Programming language --- Python 3.10.12.

\subsection{Random Sequence and Hurst Exponent}

For the short-correlated random sequences, their Hurst exponents fluctuate around the constant $0.5$ \cite{Chenjian2006}. We performed experiments with  six types random distributions, including normal distribution $\mathcal{N}(0,1)$, Chi-square distribution $\chi^2(1)$, geometric distribution $\textrm{GE}(0.25)$, Poisson distribution $\mathcal{P}(5)$, exponential distribution $\textrm{Exp}(1)$, and uniform distribution $\mathcal{U}(0,1)$, to assess the performance of our algorithms when the sample sequences are random. For each distribution, we generated $ n = 30 $ sets of sample sequences with a length of $N=10^4$ 
using the built-in random number generators in Python. Formally, the data set for the verification and validation is
\begin{equation*}
\set{X^{(\star, i)}_j: 1\le j \le 10^4}, \quad 1\le i \le n
\end{equation*}
where  $\star \in \set{\mathcal{N}(0,1), \chi^2(1), \textrm{GE}(0.25),\mathcal{P}(5), \textrm{Exp}(1),  \mathcal{U}(0,1) }$ denotes the type of distribution. Let $\hat{H}_\diamondsuit^{(\star, i)}$ be the Hurst exponent estimated from the $i$-th sample sequence $\set{X^{(\star, i)}_j: 1\le j \le 10^4}$ with the estimation method labeled by $\diamondsuit$, then we have the arithmetic average
\begin{equation*}
\mean{\hat{H}_\diamondsuit^\star} = \frac{1}{n}\sum^{n}_{i=1} \hat{H}_\diamondsuit^{(\star, i)}
\end{equation*} 
where $
\diamondsuit\in \set{\cpvar{AM}, \cpvar{AV}, \cpvar{GHE}, \cpvar{HM}, \cpvar{DFA}, \cpvar{RS}, \cpvar{TTA}, \cpvar{PM}, \cdots, \cpvar{LSSD}, \cpvar{LSV}}
$ denotes the estimation method.

For $n= 30$ times repetitive experiments, the values of $\mean{\hat{H}_\diamondsuit^\star}$ are shown in \Tab \ref{table-1} for the short-correlated random sequences. 
Note that we set the window size $ w=50 $ for calculating the optimal length $\scrd{N}{opt}$ and $ \cpvar{flag}=2 $ for the minimum $\ell_2$-norm method in linear regression.
Obviously, the Hurst exponents estimated with the thirteen algorithms mentioned above fluctuates in the range $ 0.45\sim 0.55 $, which coincides with the conclusion presented of by Chen et al \cite{Chenjian2006}.

\begin{table*}[htbp]
	\centering
	\tabcolsep=3.7pt
	\renewcommand{\baselinestretch}{1.2}
	\caption{Estimation results of Hurst exponent value for short-correlated random sequences such that $H\sim 0.5$}\label{table-1}
	\resizebox{\textwidth}{!}{
	\begin{tabular}{|c|ccccccccccccc|}
	\hline
	\diagbox{$\star$}{\rotatebox{-33}{$\mean{\hat{H}_\diamondsuit^\star}$}}{$\diamondsuit$}        
	          & AM     & AV     & GHE    & HM     & DFA    & R/S     & TTA    & PM     & AWC    & VVL    & LW     & LSSD   & LSV    \\  \hline
		$ \mathcal{N}(0,1) $   & $ 0.4942 $ & $ 0.4980 $ & $ 0.5012 $ & $ 0.4994 $ & $ 0.4919 $ & $ 0.4992 $ & $ 0.4971 $ & $ 0.5067 $ & $ 0.4971 $ & $ 0.5258 $ & $ 0.5002 $ & $ 0.4977 $ & $ 0.4987 $ \\
		$ \chi^2(1) $          & $ 0.4985 $ & $ 0.4975 $ & $ 0.4970 $ & $ 0.5234 $ & $ 0.4931 $ & $ 0.4741 $ & $ 0.5133 $ & $ 0.4992 $ & $ 0.5075 $ & $ 0.4642 $ & $ 0.4982 $ & $ 0.5001 $ & $ 0.4995 $ \\
		$ \textrm{GE}(0.25) $  & $ 0.5007 $ & $ 0.4966 $ & $ 0.5003 $ & $ 0.5155 $ & $ 0.4992 $ & $ 0.4923 $ & $ 0.5101 $ & $ 0.5105 $ & $ 0.5086 $ & $ 0.4888 $ & $ 0.5003 $ & $ 0.5009 $ & $ 0.5007 $ \\
		$  \mathcal{P}(5) $     & $ 0.4971 $ & $ 0.4931 $ & $ 0.4999 $ & $ 0.5020 $ & $ 0.4974 $ & $ 0.4966 $ & $ 0.4940 $ & $ 0.5019 $ & $ 0.4978 $ & $ 0.5071 $ & $ 0.5001 $ & $ 0.4999 $ & $ 0.5001 $ \\
		$ \textrm{Exp}(1) $    & $ 0.4837 $ & $ 0.4749 $ & $ 0.5014 $ & $ 0.5128 $ & $ 0.5030  $ & $ 0.4839 $ & $ 0.5105 $ & $ 0.4784 $ & $ 0.4946 $ & $ 0.4947 $ & $ 0.5010 $ & $ 0.5024 $ & $ 0.5021 $ \\
		$\mathcal{U}(0,1) $   & $ 0.4853 $ & $ 0.4822 $ & $ 0.5002 $ & $ 0.4896 $ & $ 0.5049 $ & $ 0.4959 $ & $ 0.4987 $ & $ 0.4975 $ & $ 0.5013 $ & $ 0.5266 $ & $ 0.5001 $ & $ 0.5022 $ & $ 0.5017 $ \\
\hline
	\end{tabular}
	}
\end{table*}

\subsection{Estimation with Fractional Gaussian Noise Sequence}\label{expr-FGN}

For the purpose of testing the accuracy of the algorithms for estimating the Hurst exponent of TS, we constructed multiple sets of experiments with the FGN sequences controlled by the Hurst exponent $ H\sim [0.3, 0.8]$. In each set of experiments, we generated $ n = 30 $ sets of sample sequences with a length of $N=3\times10^4$ via the FGN sequence generator provided by \Algr \ref{alg-fgn}. In order to reduce the variance of the estimation errors, we take the following arithmetic average
\begin{equation*}
\mean{\hat{H}_\diamondsuit^{\textrm{fgn}}} = \frac{1}{n}\sum^{n}_{i=1} \hat{H}_\diamondsuit^{(\textrm{fgn}, i)}.
\end{equation*}

For $n= 30$ times repetitive experiments, the values of $\mean{\hat{H}_\diamondsuit^{\textrm{fgn}}}$ are shown in \Tab \ref{tab-compare-alg}. The parameters $ w=50 $ and $ \cpvar{flag}=2 $
are set the same as that for \Tab \ref{table-1}. Now please recall the \Fig \ref{fig-flow} for the classification of estimation methods.
In \Tab \ref{tab-compare-alg}, it can be observed that:
\begin{itemize}
\item  The TTA method exhibits excellent accuracy in the time-domain.
\item The spectrum-domain methods are superior to the time-domain methods in general. The PM, AWC and VVL methods give similar accuracies, however the LW method is a little inferior. \\
\item For Bayesian methods, the LSSD and LSV produce the estimations of high accuracy.
\item When the sequence has long-term autocorrelation ($H>0.5$), time-domain algorithms produce underestimated values for the Hurst exponent, whereas the spectrum-domain algorithms work very well, which is consistent with the results discovered by Chen et al.   \cite{Chenjian2006}.
\end{itemize}

\begin{table*}[htbp]
	\centering
	\renewcommand{\baselinestretch}{1.2}
	\tabcolsep=5pt
	\caption{Comparison of estimation accuracy of the 13 algorithms via  FGN sequences with given Hurst exponent}\label{tab-compare-alg}
	\resizebox{\textwidth}{!}{
	\begin{tabular}{|c|ccccccc|cccc|cc|}
\hline
\diagbox{$\scru{H}{fgn}$}{\rotatebox{-33}{$\mean{\hat{H}_\diamondsuit^\textrm{fgn}}$}}{$\diamondsuit$}    & AM     & AV     & GHE    & HM     & DFA    & \fbox{R/S}     & \underline{TTA}    & \underline{PM}     & AWC    & VVL    & LW     & \underline{LSSD}   & \underline{LSV}    \\ \hline
		$0.30 $ & $ 0.3023 $ & $ 0.2984 $ & $ 0.3006 $ & $ 0.3006 $ & $ 0.3078 $ & $ 0.3692 $ & $ 0.3003 $ & $ 0.3099 $ & $ 0.2919 $ & $ 0.3126 $ & $ 0.2629 $ & $ 0.3002 $ & $ 0.3003 $ \\
		$0.35 $ & $ 0.3521 $ & $ 0.3495 $ & $ 0.3501 $ & $ 0.3500 $ & $ 0.3502 $ & $ 0.4056 $ & $ 0.3500 $ & $ 0.3614 $ & $ 0.3426 $ & $ 0.3592 $ & $ 0.3239 $ & $ 0.3496 $ & $ 0.3497 $ \\
		$0.40 $ & $ 0.4074 $ & $ 0.4044 $ & $ 0.3994 $ & $ 0.3999 $ & $ 0.3981 $ & $ 0.4468 $ & $ 0.3988 $ & $ 0.4047 $ & $ 0.3952 $ & $ 0.4085 $ & $ 0.3844 $ & $ 0.4001 $ & $ 0.3998 $ \\
		$0.45 $ & $ 0.4365 $ & $ 0.4353 $ & $ 0.4507 $ & $ 0.4511 $ & $ 0.4544 $ & $ 0.4856 $ & $ 0.4472 $ & $ 0.4369 $ & $ 0.4372 $ & $ 0.4402 $ & $ 0.4429 $ & $ 0.4503 $ & $ 0.4504 $ \\
		$0.50 $ & $ 0.5000 $ & $ 0.4956 $ & $ 0.4991 $ & $ 0.4991 $ & $ 0.4995 $ & $ 0.5293 $ & $ 0.4994 $ & $ 0.5037 $ & $ 0.4989 $ & $ 0.4993 $ & $ 0.4994 $ & $ 0.4983 $ & $ 0.4985 $ \\
		$0.55 $ & $ 0.5425 $ & $ 0.5397 $ & $ 0.5511 $ & $ 0.5512 $ & $ 0.5516 $ & $ 0.5702 $ & $ 0.5510 $ & $ 0.5561 $ & $ 0.5431 $ & $ 0.5430 $ & $ 0.5577 $ & $ 0.5521 $ & $ 0.5519 $ \\
		$0.60 $ & $ 0.5869 $ & $ 0.5849 $ & $ 0.5999 $ & $ 0.6001 $ & $ 0.5991 $ & $ 0.6102 $ & $ 0.6017 $ & $ 0.5851 $ & $ 0.5985 $ & $ 0.6005 $ & $ 0.6127 $ & $ 0.6000 $ & $ 0.6003 $ \\
		$0.65 $ & $ 0.6411 $ & $ 0.6394 $ & $ 0.6478 $ & $ 0.6474 $ & $ 0.6485 $ & $ 0.6539 $ & $ 0.6482 $ & $ 0.6524 $ & $ 0.6491 $ & $ 0.6432 $ & $ 0.6652 $ & $ 0.6492 $ & $ 0.6488 $ \\
		$0.70 $ & $ 0.6679 $ & $ 0.6648 $ & $ 0.6984 $ & $ 0.6988 $ & $ 0.7057 $ & $ 0.6853 $ & $ 0.6994 $ & $ 0.6824 $ & $ 0.6955 $ & $ 0.6964 $ & $ 0.7213 $ & $ 0.7002 $ & $ 0.7000 $ \\
		$0.75 $ & $ 0.7192 $ & $ 0.7169 $ & $ 0.7470 $ & $ 0.7472 $ & $ 0.7516 $ & $ 0.7226 $ & $ 0.7484 $ & $ 0.7459 $ & $ 0.7519 $ & $ 0.7447 $ & $ 0.7747 $ & $ 0.7496 $ & $ 0.7496 $ \\
		$0.80 $ & $ 0.7636 $ & $ 0.7635 $ & $ 0.7948 $ & $ 0.7948 $ & $ 0.7978 $ & $ 0.7551 $ & $ 0.7990 $ & $ 0.7955 $ & $ 0.8052 $ & $ 0.7998 $ & $ 0.8288 $ & $ 0.7997 $ & $ 0.8002 $ \\
\hline
	\end{tabular}
	}
\end{table*}

\subsection{Relative Error of Estimating Hurst Exponent}

\label{subsec-relative-error}

With the help of the controllable Hurst exponent $\scru{H}{fgn}$ for the FGN sequences, we can compare  the difference of various estimation methods mentioned above.  

For the sequence $\set{X_j: 1\le j \le N}$ generated from the FGN sequences, suppose the estimated Hurst exponent in the $i$-th experiment in $n$ repeatable experiments with estimation method $\diamondsuit$ is  $\scrd{\hat{H}}{\diamondsuit}^i$. The relative error of estimation can be defined by
\begin{equation}\label{rela-error}
	\scrd{\eta}{\diamondsuit}^i = \dfrac{\abs{\scrd{\hat{H}}{\diamondsuit}^i-
	\scru{H}{fgn}}}{\scru{H}{fgn}} \times 100 \%, \quad 1\le i\le n
\end{equation}
then the average relative error must be 
\begin{equation} \label{rela-error-ave}
\scrd{\eta}{\diamondsuit} = \frac{1}{n}\sum^{n}_{i=1} \scrd{\eta}{\diamondsuit}^i. 
\end{equation}
In consequence, for different estimation $\diamondsuit$, we can compare their relative error to evaluate their performances with the help of \eqref{rela-error} and \eqref{rela-error-ave}.

\begin{figure*}[htb]
\centering
\subfigure[Relative Error for Time-Domain Estimators]{
		\includegraphics[width=0.45\textwidth]{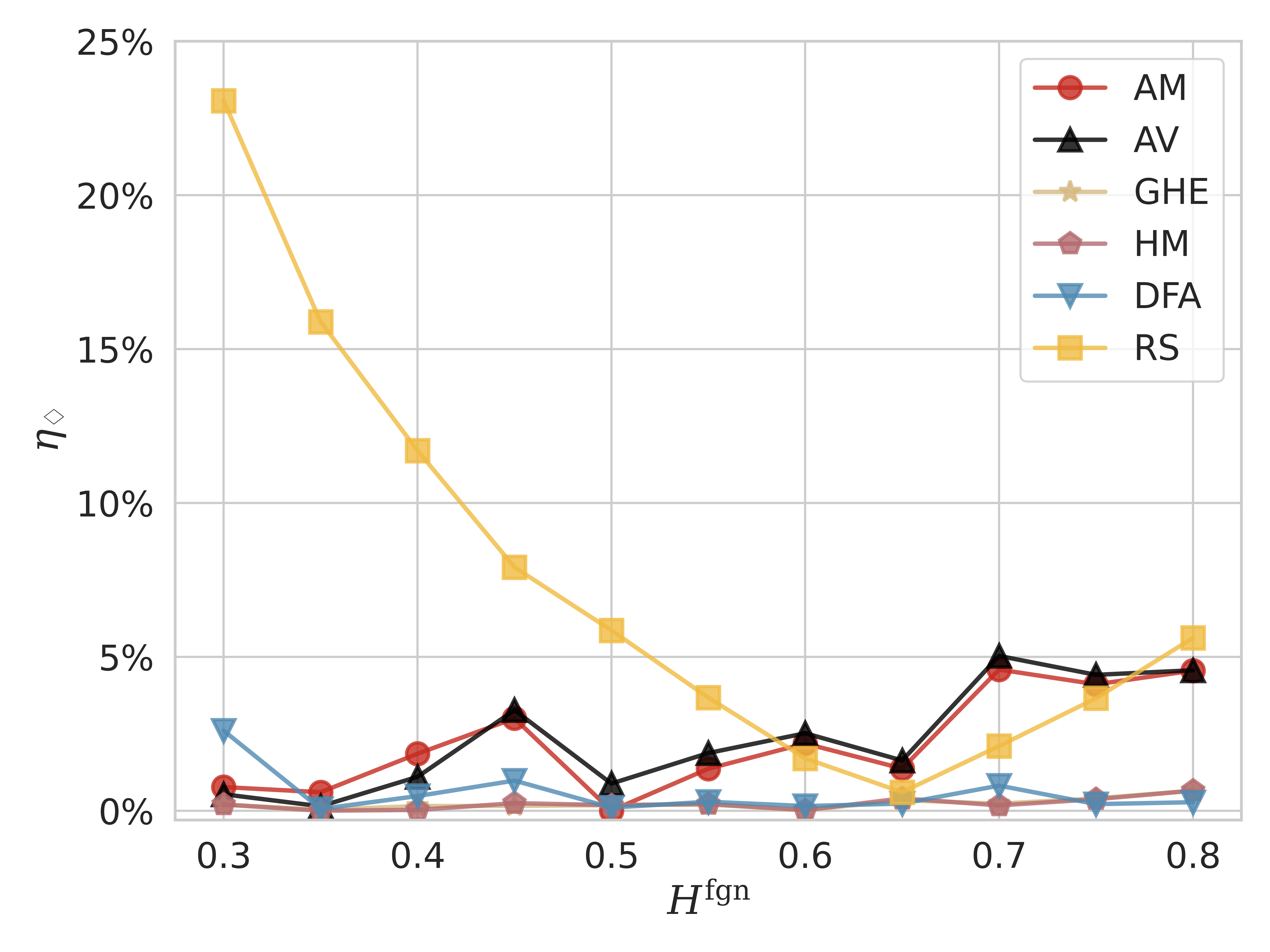}
		\label{fig-relative-err-td}
}
\subfigure[Relative Error for Spectrum-Domain Estimators]{
		\includegraphics[width=0.45\textwidth]{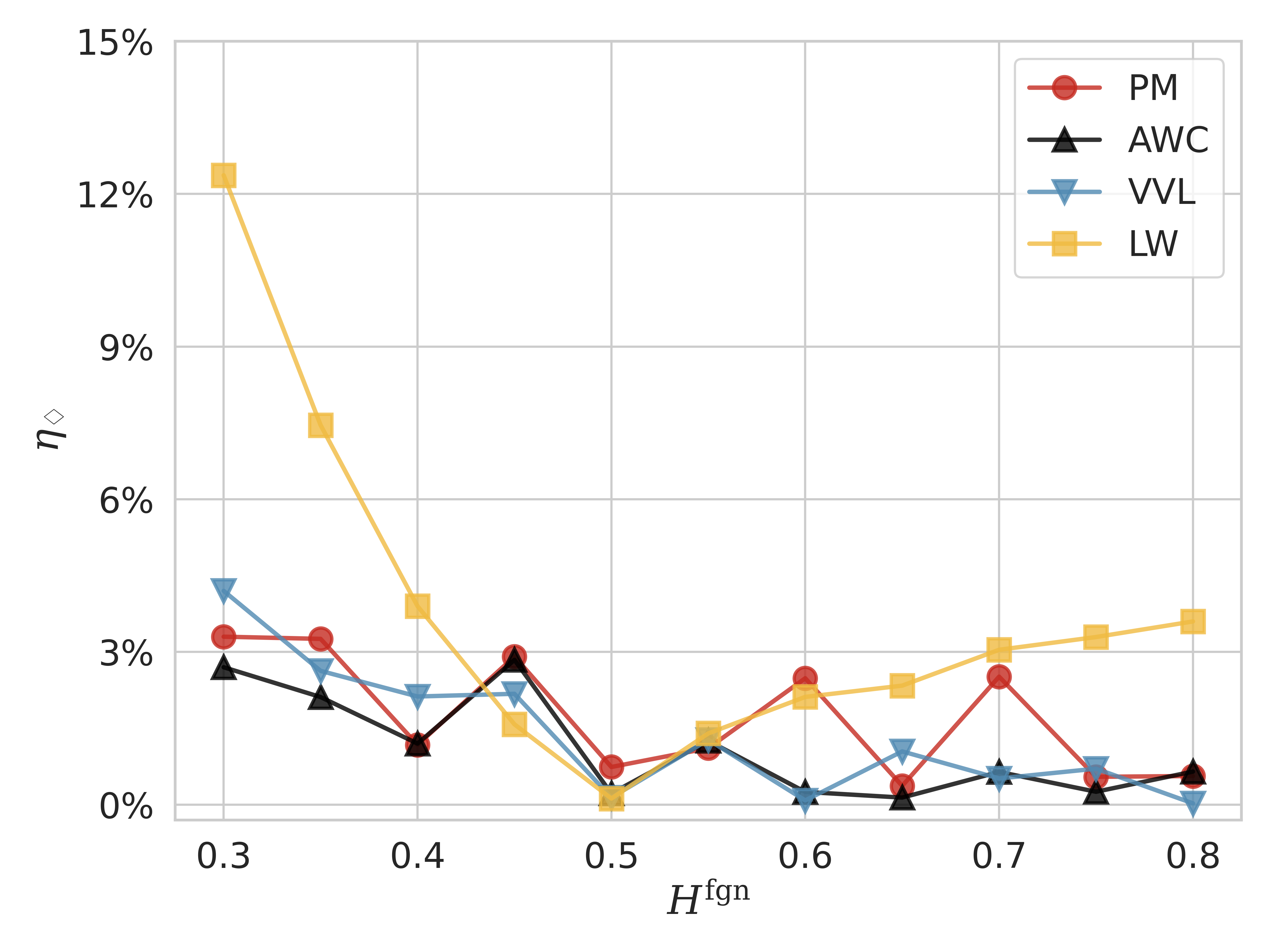}
		\label{fig-relative-err-fd}
}
\caption{Comparison of Hurst Estimators via the FGN sequences with known Hurst exponent}
\label{figure-3}
\end{figure*}

\Fig \ref{figure-3} illustrates the values of $\eta_\diamondsuit$ in $n=100$ repeatable experiments with various estimation methods for the Hurst exponent of the FGN sequences. The results of the time-domain methods and the spectrum-domain methods are displayed in \Fig \ref{fig-relative-err-td} and \Fig \ref{fig-relative-err-fd} respectively. Please note that the relative errors for the two Bayesian statistical methods (LSSD and LSV) and the TTA-method are not shown in the figure due to their small values (about $1.5\times 10^{-3}$).

In the \Fig \ref{fig-relative-err-td} for the time-domain methods, as the Hurst exponent value of the FGN sample sequence increases, the estimation error of most methods exhibits an upward trend. On the contrary, the relative error for the R/S method decreases with the nominal value of Hurst exponent. An interesting phenomena is that the relative error of the DFA method remains stable and its value  is smaller than $5\times10^{-3}$.  

In the \Fig \ref{fig-relative-err-fd} for the spectrum-domain methods, all methods achieve relatively accurate estimation results. However, the error curve of the LW method exhibits a symmetric distribution around $H=0.5$. To address the issue of larger errors in the lower range of $ \scru{H}{fgn} $ for the R/S method, we can improve the estimation precision by constructing the revised statistical $ \mathscr{R}_{Y}^{\mathrm{AL}}$ as given in equation \eqref{RS-6}.

\subsection{Impacts of Norm and Optimization Method on Estimation Performance}

\begin{figure*}[htb]
\centering
\subfigure[$\scru{H}{fgn}=0.3 $]{
		\includegraphics[width=0.3\textwidth]{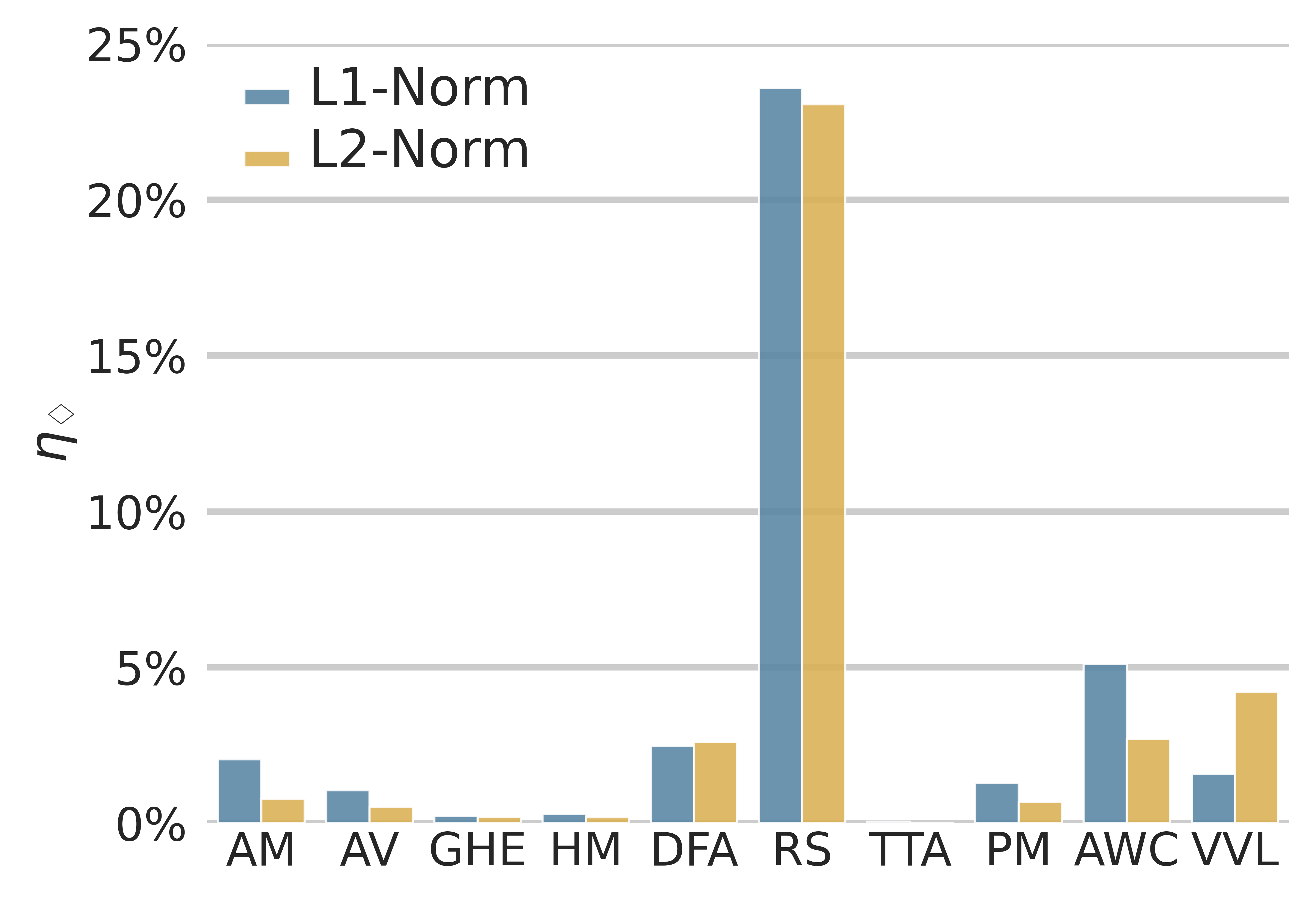}
		\label{fig-relative-err-norm-0.3}
}
\subfigure[$ \scru{H}{fgn}=0.5 $]{
		\includegraphics[width=0.3\textwidth]{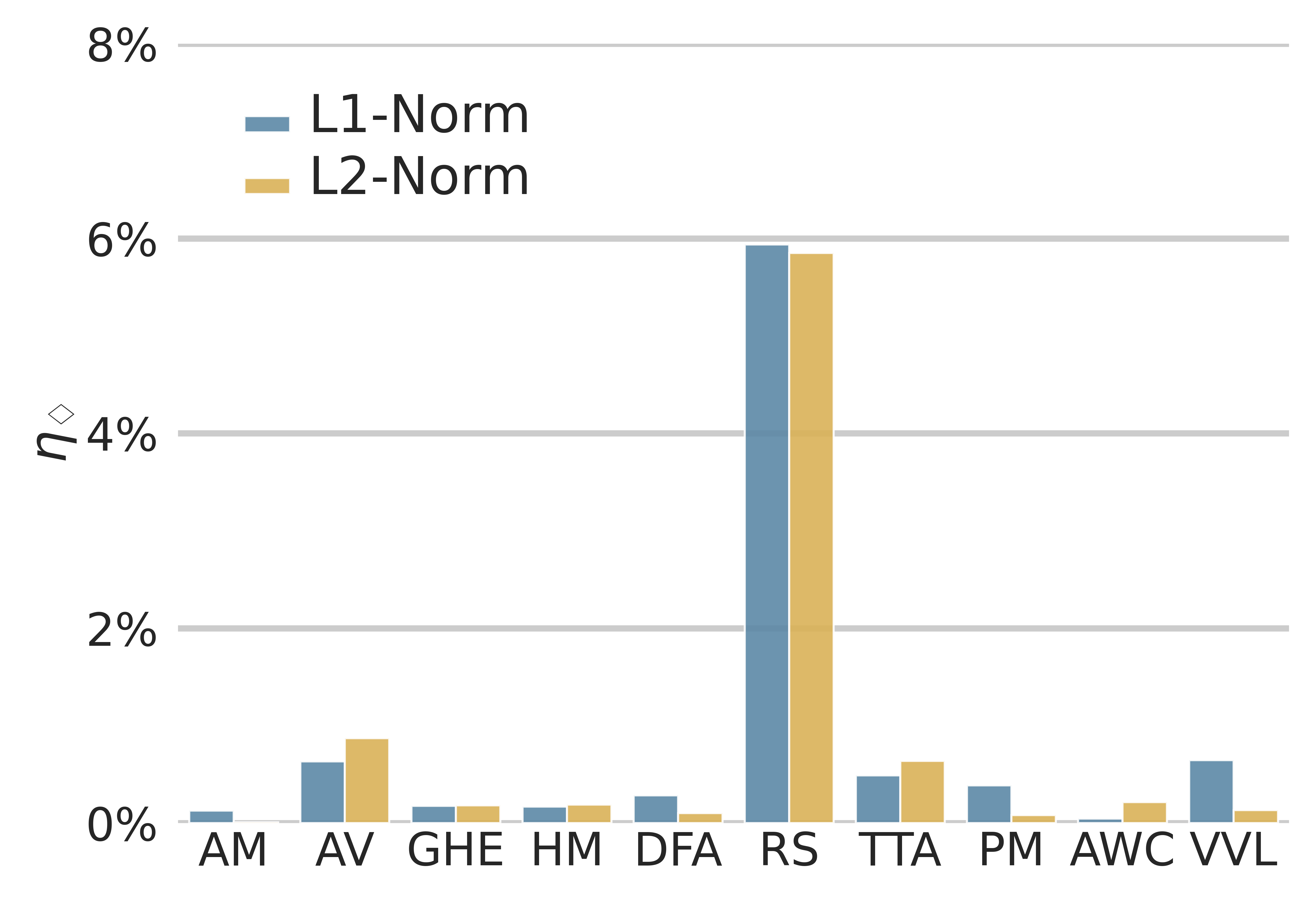}
		\label{fig-relative-err-norm-0.5}
}
\subfigure[$ \scru{H}{fgn}=0.8 $]{
		\includegraphics[width=0.3\textwidth]{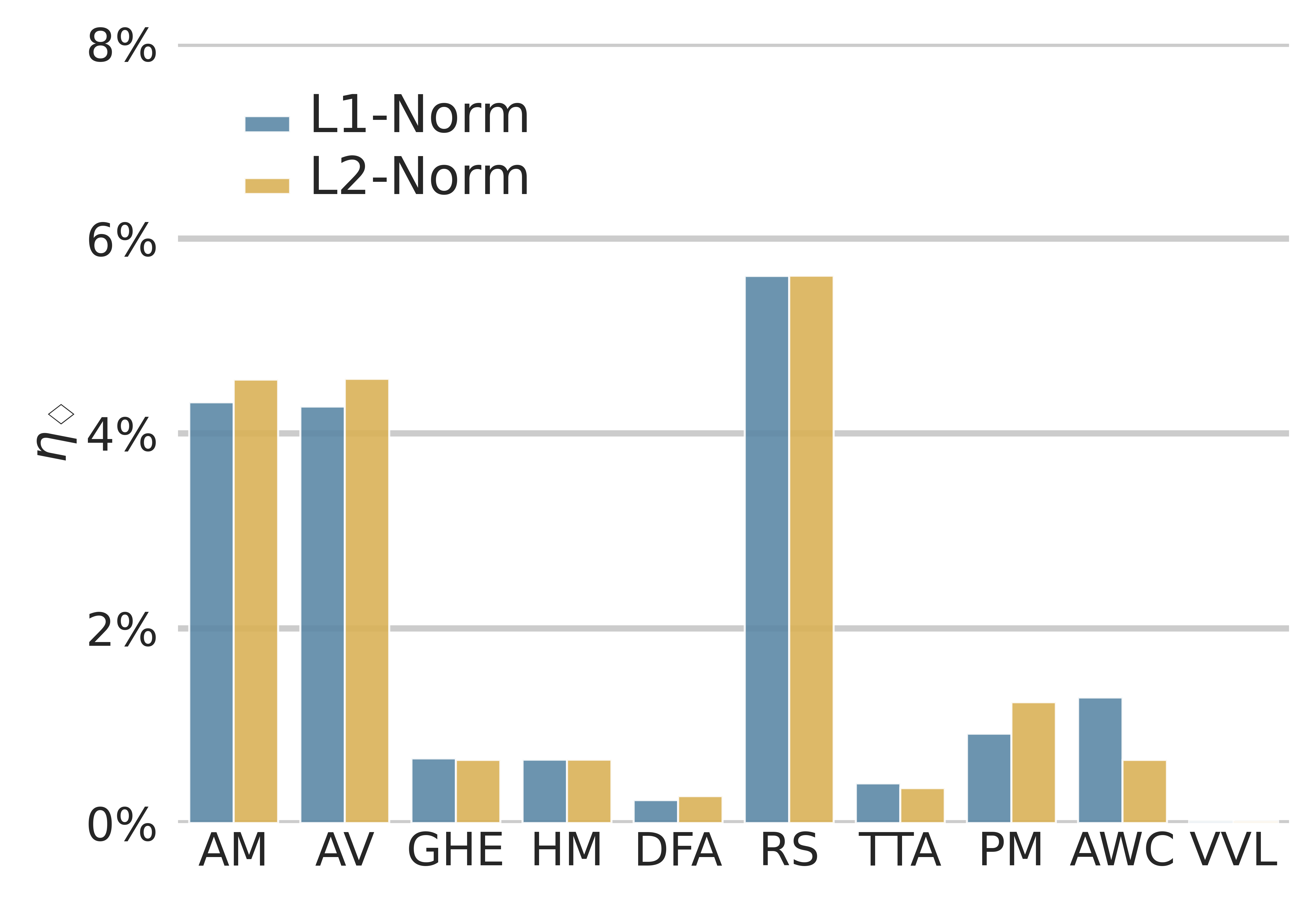}
		\label{fig-relative-err-norm-0.8}
}
	\caption{Relative Errors for two linear regression methods}\label{L1-vs-L2-figure-1}
\end{figure*}

In the experiment  mentioned in sub-section \ref{expr-FGN}, we also explored the application of different linear regression methods in the estimation of the Hurst exponent. For example, in three sets of experiments with $\scru{H}{fgn}\in \set{0.3, 0.5, 0.8}$, we selected the $ \ell_1 $-norm and $ \ell_2 $-norm as the optimization methods for linear regression. We fitted the parameters $ \mat{A},\vec{b} $ obtained by each algorithm and calculated the relative errors by \eqref{rela-error}, the results as shown in Figure \ref{L1-vs-L2-figure-1}:
\begin{itemize}
\item Figure \ref{fig-relative-err-norm-0.3}, Figure \ref{fig-relative-err-norm-0.5}
and Figure \ref{fig-relative-err-norm-0.8} show that the average relative error $\scrd{\eta}{\diamondsuit}$ is large than $5\%$ just for the RS method,  which does not depend on the optimization method for parameter estimation. Moreover, the estimation precision for GHE and HM methods are the same, which also does not depend on the usage of $\ell_1$-norm and $\ell_2$-norm optimization. 
\item Figure \ref{fig-relative-err-norm-0.3} shows that when $\scru{H}{fgn} $ is small, the AWC and VVL methods are sensitive for the usage of $\ell_1$-norm and $\ell_2$-norm optimization.
\item Figure \ref{fig-relative-err-norm-0.5} shows that when $\scru{H}{fgn} $ is around $0.5$, only the VVL method is sensitive for the usage of $\ell_1$-norm and $\ell_2$-norm optimization.
\item Figure \ref{fig-relative-err-norm-0.8} shows that when $\scru{H}{fgn} $ is large, only the AWC method is sensitive for the usage of $\ell_1$-norm and $\ell_2$-norm optimization.
\end{itemize}

It can be observed that most algorithms for estimating the Hurst exponent have relative estimation errors controlled at below $ 5\% $ when the length of the TS is long enough. In general, the choice off $\ell_1$-norm and $\ell_2$-norm optimization is not essential for the long sequences. When the length of the TS is short, we recommend the minimum $\ell_1$-norm optimization for parameter estimation.

\subsection{Hurst Exponent of Reaction Time Sequence}

\begin{figure}[htbp]
	\centering
	\includegraphics[width=.5\textwidth]{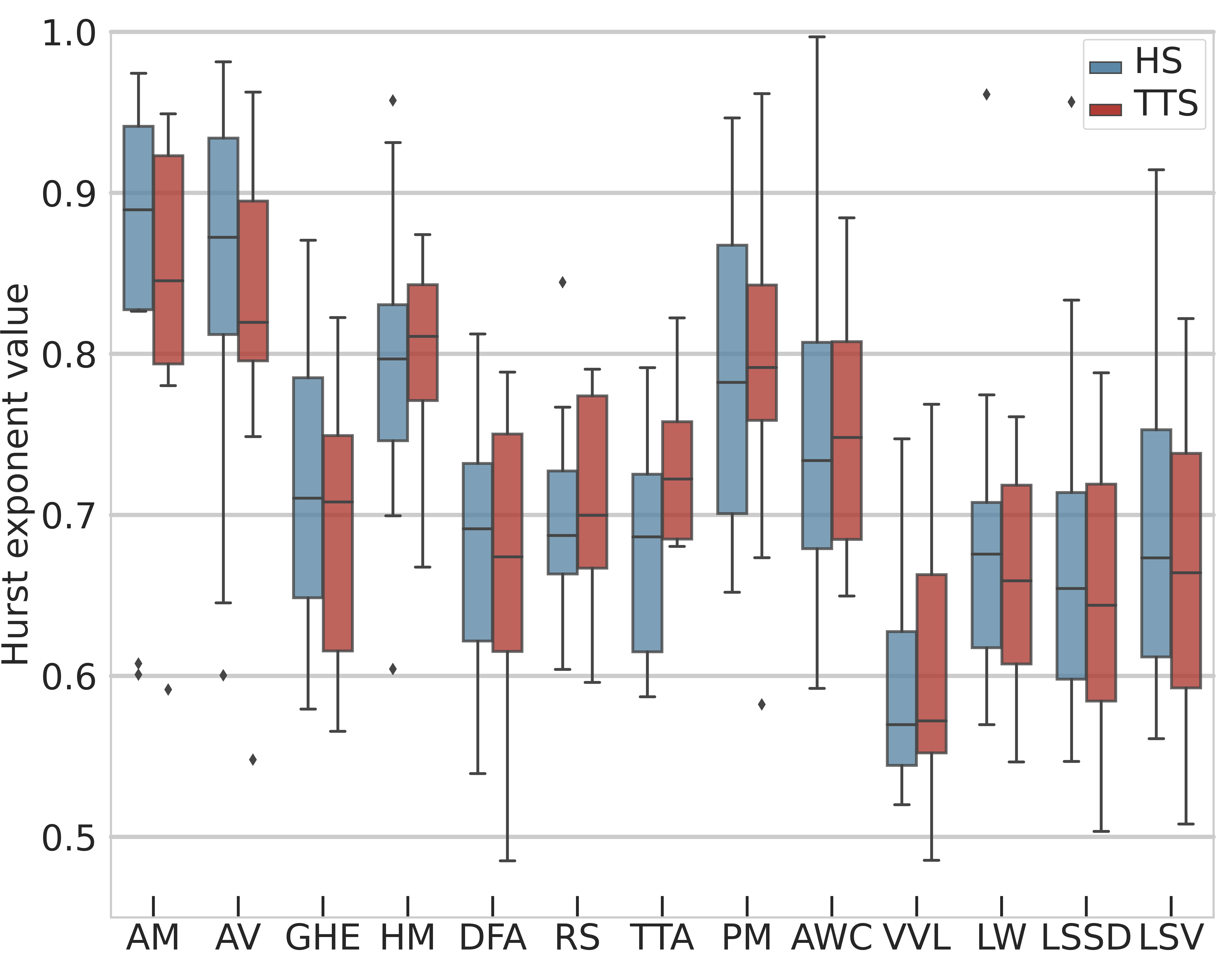}
	\caption{Estimation of reaction time sequence in HS-test and TTS-test by each method}\label{figure-2}
\end{figure}

In the study conducted by Lauren et al.  in 2021  \cite{Bloomfield2021perceiving}, a series of speech testing experiments were designed, which included the \textit{human-speaker} (HS) test and the \textit{text-to-speech} (TTS) test. The \textit{reaction time} (RT) data from 20 participants in both test groups were collected and made publicly accessible for retrieval\footnote{\url{https://royalsocietypublishing.org/doi/suppl/10.1098/rsif.2021.0272}}. In 2023, Likens applied two Bayesian methods and the DFA method to estimate the Hurst exponent of the reaction time data   \cite{Likens2023better}. The results showed that all these reaction time sequences exhibited long-range memory characteristics $(H>0.5)$. In this study, we also employed these data to evaluate the accuracy of the $13$ methods discussed above (with a window size parameter set to $w = 50$ and 
the $\ell_2$-norm optimization for the linear regression). The experimental results are illustrated in \Fig \ref{figure-2}.

In \Fig \ref{figure-2}, the box-body illustrates the Hurst exponent values of the reaction time sequences exhibited by 20 experimental subjects in different tests under the estimation. It is obvious that both the HS-test data set and the TTS-test data set demonstrate long-term memory characteristics in the estimated results$(H>0.5)$. This experimental result aligns with the conclusion obtained by Zhang et al. \cite{Zhang2017r} in 2017.

\subsection{Discussion}

The results from subsection \ref{subsec-relative-error} indicates that the estimation accuracy of spectrum-domain methods is significantly superior to that of time-domain methods, with the two Bayesian methods demonstrating the highest precision. In terms of method selection strategy, we have the following observations:
\begin{itemize}
	\item Time-domain methods exhibit good interpretation and  no advanced programming skills are needed for implementing the estimation algorithms, whereas the implementation process of spectrum-domain methods relies on more advanced mathematical tools such as the FFT and wavelet analysis.
	\item With the time-domain methods, we can effectively demonstrate the correspondence between partial statistical properties of sequences and sample scales (typically represented by a straight line). This is also why time-domain methods are highly popular and widely applied.
	\item The Bayesian methods provides good accuracy, but the optimization algorithms or fixed-point algorithms are necessary. 
\end{itemize}  

In addition to the 13 estimation methods mentioned above, there are other methods not mentioned in this paper. For example, the \textit{maximum likelihood estimation} (MLE) method for estimating the Hurst exponent is known for its implementation difficulty and high time complexity. For the interested readers, please refer to the references  \cite{Guerrero2005maximum, chang2014efficiently, garcin2022comparison}.

In 2022, G{\'o}mez et al. proposed the \textit{Kolmogorov-Smirnov} (KS) method based on the GHE method and TTA method  \cite{Gomez2022improvement}. The KS method estimates the Hurst exponent by calculating the Kolmogorov-Smirnov (KS) statistic distance between the empirical distributions of samples  \cite{Hodges1958significance}.  Gomez et al. \cite{Gomez2022improvement} provided the Python code for the KS method and it is omitted here. 

In our experiment, it has been observed that for certain shorter time sequences   \cite{Li2018systematic}, most estimation methods fail to produce accurate results. However, there exist a few methods that have good estimation performance, such as the GHE method, LW method, and LSVmethod. As for other time-domain methods, such as the R/S method, we can take the linear interpolation technique and use interpolation points to construct the sample sequences of lengths $ \set{N/2, N/4, \cdots} $. This approach extends the applicability of the estimation methods for the Hurst exponent to the shorter time sequences.

It should remarked that there are also possible ways for estimating the Hurst exponent via the new techniques for analyzing the time sequences with LTM. For the  time-domain algorithms, the TTA method has demonstrated its commendable performance. The topological structures of the time sequences \cite{GIDEA2018TS-TDA,KARAN2021TS-TDA,ElYaagoubi2023TS-TDA} are attracting more and more researchers to discover more intrinsic characteristics. For the frequency-domain methods, the two wavelet-based approaches, viz. AWC and VVL,  have demonstrated novel perspectives for us. The mode decomposition of time series \cite{Shang2019TS-EMD,PAN2019189SGMD} is a new interesting point of current researches, and this may potentially lead to the development of additional computational methods in the frequency-domain.

\section{Conclusions}

\label{sec-conclusions}

In this paper, we summarized 13 typical methods for estimating the Hurst exponent and categorized them into different classes based on different strategies:
\begin{itemize}
	\item time-domain methods and spectrum-domain methods based on the the representation of time sequence;
	\item linear regression methods and Bayesian methods based on the parameter estimation method.
\end{itemize}
Both the mathematical principle and algorithmic pseudo-codes are provided for these 13 methods, which helps the researchers and potential users to implement these methods with concrete programming language based on a unified framework.

Our contributions are summarized as follows:
\begin{itemize}
\item A general sequence partition  framework was proposed for various time-domain methods based on the optimal approximate length and feasible sequence grouping approach.
\item The fixed-point algorithm, local minimum search algorithm,  and linear regression method based on $\ell_1$-norm are applied to improve the accuracy of estimating the Hurst exponent with available estimation methods.
\item  The estimation methods are classified with two perspectives, viz. the sequence representation and parameter estimation. 
\item The sequences generated via FGN and pure random sequences are used to design a series of experiments to test the accuracy of the 13 estimators discussed above. 
\item The flowcharts of R/S method and DFA method are provided for helping the readers to understand the essence and steps of the algorithms concerned.
\end{itemize}

The numerical experiments and error analysis for the 13 estimation methods shows that:
\begin{itemize}
\item The estimation accuracy of spectrum-domain methods is superior to time-domain methods, with a relative error of less than $6\%$ in general.
\item When the value of Hurst exponent is small (say $H < 0.35$), the relative errors of the estimation obtained by the R/S method and LW method are significantly larger than $5\%$;
\item The choice of $\ell_1$-norm and $\ell_2$-norm has little impact on the estimation accuracy.  
\item The estimation with the practical data captured from the human behavioral experiment available online implies that  each estimation method can effectively reveal the long-term memory features of the sequences, which confirming the suitability of the 13 methods.
\end{itemize}

For the off-line applications where the Hurst exponent in involved, we recommend the TTA method for the time-domain methods, the LSSD and LSV for the Bayesian methods, and the PM method for the spectrum-domain methods. For the real-time applications in which the Hurst exponent should be estimated dynamically, we recommend the R/S method to reduce the computational complexity since both the range and standard deviation  can be estimated iteratively with the time clock.

It should be noted that for the online applications there is no acceptable algorithm for estimating the Hurst exponent currently. It is our future work to design the real-time algorithm for estimating the Hurst exponent in an iterative way.

\section*{Acknowledgments}

This work was supported  by the Research Project on Education and Teaching Reform in Higher Education System of Hainan Province under grant numbers Hnjg2019-50 and Hnjg2023ZD-26, in part by the National Natural Science Foundation of China under grant number
62167003, in part by the Hainan Provincial Natural Science Foundation of China under grant numbers 720RC616 and 623RC480, and in part by the Specific Research Fund of The Innovation Platform for
Academicians of Hainan Province.

\section*{Code Availability}

The code for the implementations of the algorithms discussed in this paper can be downloaded from the following GitHub website
\begin{center}
\textcolor{blue}{\href{https://github.com/GrAbsRD/HurstExponent}{https://github.com/GrAbsRD/HurstExponent}}
\end{center}
For the purpose of easy usage, both the Python and the Octave/MATLAB codes are provided.

\section*{Data Availability}

The data set supporting the results of this study is available on the GitHub website
\begin{center}
	\textcolor{blue}{\href{https://github.com/ZhiQiangFeng/HurstExponentTestData}{https://github.com/ZhiQiangFeng/HurstExponentTestData}}
\end{center}
The repository includes a README file with essential information on data access and use. For any questions regarding the data, please contact the author with the e-mail: \verb|z.q.feng@foxmail.com|.

\section*{Conflict of Interest Statement}

The authors declare that they have no known competing financial interests or personal relationships that could have appeared to influence the work reported in this paper.

\section*{How to Cite This Work?}

\fbox{
\begin{minipage}{16cm}
\noindent  H. -Y. Zhang, Z. -Q. Feng, S. -Y. Feng and Y. Zhou. Typical Algorithms for Estimating Hurst Exponent of Time Sequence: A Data Analyst’s Perspective, \textit{IEEE Access}, 12(12): 185528--185556, 2024, doi: 10.1109/ACCESS.2024.3512542. arXiv: 2310.19051v4
\end{minipage}
}

\end{document}